\documentclass[paper]{JHEP3} 

\usepackage{epsfig,multicol,multirow}
\usepackage{amsmath,subfigure,latexsym,amssymb}
\usepackage{here}

\newcommand\fverb{\setbox\fverbbox=\hbox\bgroup\verb}
\newcommand\fverbdo{\egroup\medskip\noindent%
			\fbox{\unhbox\fverbbox}\ }
\newcommand\fverbit{\egroup\item[\fbox{\unhbox\fverbbox}]}
\newbox\fverbbox

\def\lsi{\raise0.3ex\hbox{$<$\kern-0.75em\raise-1.1ex\hbox{$\sim$}}}
\def\gsi{\raise0.3ex\hbox{$>$\kern-0.75em\raise-1.1ex\hbox{$\sim$}}}

\newcommand{\beq}{\begin{equation}}
\newcommand{\eeq}{\end{equation}}
\newcommand{\beqa}{\begin{eqnarray}}
\newcommand{\eeqa}{\end{eqnarray}}

\allowdisplaybreaks

\title{Direct test of the AdS/CFT correspondence
by Monte Carlo studies of ${\cal N}=4$ super Yang-Mills theory}
\author{Masazumi Honda,$^{a,b,c}$ Goro Ishiki,$^{b,c,d}$
Sang-Woo Kim,$^{b,d,e,f}$
Jun Nishimura$^{a,b}$ \hspace{3cm} and Asato Tsuchiya$^{g}$
\vspace*{0.3cm} \\
\llap{$^a$}Department of Particle and Nuclear Physics,\\
Graduate University for Advanced Studies (SOKENDAI),\\
Tsukuba, Ibaraki 305-0801, Japan\\
\llap{$^b$}High Energy Accelerator Research Organization (KEK),\\
Tsukuba, Ibaraki 305-0801, Japan\\
\llap{$^c$}Yukawa Institute for Theoretical Physics, Kyoto University,\\
Kitashirakawa Oiwakecho, Sakyo-ku, Kyoto 606-8502, Japan\\
\llap{$^d$}Center for Quantum Spacetime (CQUeST),\\
Sogang University, Seoul 121-742, Korea\\
\llap{$^e$}Department of Physics, Osaka University,\\
Toyonaka, Osaka 560-0043, Japan\\
\llap{$^f$}School of Physics, Korea Institute for Advanced Study,\\
85 Hoegiro Dongdaemun-gu, Seoul 130-722, Korea\\
\llap{$^g$}Department of Physics, Shizuoka University,\\
836 Ohya, Suruga-ku, Shizuoka 422-8529, Japan
\vspace*{0.3cm} \\
\email{mhonda@post.kek.jp, ishiki@post.kek.jp, sang@kias.re.kr,\\
jnishi@post.kek.jp, satsuch@ipc.shizuoka.ac.jp}}

\preprint{KEK-TH-1646,~YITP-13-65,~KIAS-P13030}

\abstract{
We perform nonperturbative studies of
${\cal N}=4$ super Yang-Mills theory
by Monte Carlo simulation.
In particular,
we calculate
the correlation functions of chiral primary operators
to test the AdS/CFT correspondence.
Our results agree with the predictions obtained from
the AdS side that
the SUSY non-renormalization property
is obeyed by the three-point functions but \emph{not}
by the four-point functions
investigated in this paper.
Instead of the lattice regularization, we
use a novel regularization of the theory
based on an equivalence in the large-$N$ limit
between
the ${\cal N}=4$ SU($N$) theory on $R \times S^3$
and a one-dimensional SU($N$) gauge theory known as
the plane-wave (BMN) matrix model.
The equivalence extends
the idea of large-$N$ reduction
to a curved space and, at the same time, overcomes the obstacle
related to the center symmetry breaking.
The adopted regularization
for $S^3$ preserves
16 SUSY, which is crucial
in testing the AdS/CFT correspondence with the
available computer resources.
The only SUSY breaking effects, which come from the
momentum cutoff $\Lambda$ in $R$ direction, are made negligible
by using sufficiently large $\Lambda$.
}


\begin{document}

\section{Introduction}

Supersymmetry (SUSY) is a symmetry, which is expected to play a key
role in understanding particle physics beyond the electroweak scale
up to the Planck scale.
On one hand, it provides
a natural way to stabilize the electroweak scale
against quantum corrections
from the viewpoint of a fundamental theory anticipated
to appear at the Planck scale.
On the other hand, SUSY is crucial in various new approaches to
superstring theory and quantum gravity based on gauge theories.
In either of these applications, it is important to study
supersymmetric gauge theories
in a complete nonperturbative formulation such as the lattice gauge theory.
This is considered very hard, however, due to the fact
that the lattice inevitably breaks SUSY.
Namely the SUSY algebra includes
translational symmetry, which is manifestly broken down
to a discrete one by the lattice regularization.
In order to restore SUSY in the continuum limit,
one typically needs
to fine-tune the parameters in the lattice action.
Recently various ideas for preserving SUSY as much as possible
in lattice and non-lattice regularizations
have been put forward.
%

In this paper we
perform Monte Carlo studies of
${\cal N}=4$ super Yang-Mills theory (SYM),
which is a four-dimensional superconformal
gauge theory with maximal 32 supersymmetries.
%
In addition to the superconformality and maximal supersymmetry
in four dimensions,
the theory has attracted much attention for
the conjectured Montonen-Olive duality,
which realizes the strong-weak duality as an extension
of the electromagnetic duality of Maxwell's theory.
In the context of superstring theory,
the ${\cal N}=4$ U($N$) SYM appears
as the low-energy effective theory of a stack of $N$ D3-branes.
This fact led to the famous
AdS/CFT correspondence \cite{AdS-CFT,Aharony:1999ti,%
Gubser:1998bc,Witten:1998qj},
which is a conjectured duality relation between
the ${\cal N}=4$ SU($N$) SYM and
type IIB superstring theory
on $AdS_5\times S^5$.
In the $N\rightarrow \infty$ limit with fixed but large
't Hooft coupling $\lambda \equiv g_{\rm YM}^2 N$, in particular,
the string theory side reduces to the classical supergravity,
which enables
various explicit calculations.
The AdS/CFT correspondence thus provided
a lot of predictions on the strong coupling dynamics of
${\cal N}=4$ SU($N$) SYM
albeit in the large-$N$ limit.
Some of the predictions have been confirmed in a remarkable manner
assuming integrability \cite{spin-chain},
and there are also some attempts to prove the AdS/CFT correspondence
based on the worldsheet approach \cite{kawai-suyama,Berkovits:2007rj}.
However, a direct test based on first-principle calculations in
${\cal N}=4$ SU($N$) SYM would still be very important.

In the case of D0-branes, one obtains the gauge/gravity duality
between one-dimensional SUSY gauge theory with 16 supercharges
and the black 0-brane solution in type IIA supergravity \cite{Itzhaki:1998dd}.
Monte Carlo studies of the 1d SUSY gauge theory confirmed
various predictions from the gauge/gravity
duality \cite{Hanada:2007ti,%
Catterall:2007fp,AHNT,Catterall:2008yz,%
Hanada:2008gy,Hanada:2008ez,Catterall:2009xn,Hanada:2009ne,%
Nishimura:2009xm,Hanada:2011fq,Nishimura:2012xs,Kadoh:2012bg}.
In particular,
the black hole thermodynamics
has been reproduced \cite{Hanada:2008ez}
{\it including $\alpha'$ corrections},
which correspond to the effects of closed strings with finite length.
Predictions for Wilson loops and correlation functions
have also been confirmed \cite{Hanada:2008gy,Hanada:2009ne,Hanada:2011fq}.
The success of these works largely depends on the fact that
the 1d gauge theory does not have UV divergences,
and hence all the 16 SUSYs can be restored without fine-tuning.

Monte Carlo studies of SUSY gauge theories in more than two
dimensions require more ideas.
For instance, refs.~\cite{Hiller:2000nf,Hiller:2005vf}
studied 2d SYM
in the large-$N$ limit
numerically by using the discrete light-cone quantization.
The two-point correlation function of the stress-energy tensor
has been calculated, and the expected power-law behavior has been confirmed.
Refs.~\cite{Hanada:2012si,Honda:2012bx} studied
the ABJM theory, which is
a 3d superconformal Chern-Simons gauge theory describing M2-branes,
by simulating a matrix model which can be obtained
by applying the so-called localization technique \cite{Pestun:2007rz}
to the 3d theory.
See also refs.~\cite{Kikukawa:2008xw,Catterall:2009it,Joseph:2011xy,D'Adda:2009kj,%
Hanada:2009hq,Catterall:2010fx,Catterall:2010gf,%
Hanada:2011qx,Kanamori:2012et,Takimi:2012zw,%
Catterall:2012yq}
for recent works on lattice formulations of SUSY theories.
%
As for 4d ${\cal N}=4$ SYM,
any lattice formulations proposed so far
seem to require fine-tuning of at least
three parameters \cite{Kaplan:2005ta,Unsal:2005us,%
Elliott:2008jp,Catterall:2008dv,Giedt:2009yd}.
(See refs.~\cite{Catterall:2011pd,Catterall:2013roa}, however.)
The non-lattice regularization \cite{Ishii:2008ib}
used in this paper
respects 16 out of 32 SUSYs of the theory,
and it avoids the fine-tuning problem completely.
The idea
is based on
the large-$N$ reduction and fuzzy spheres, which reduce
4d $\mathcal{N}=4$ SYM to
1d SUSY gauge theory with certain
deformation preserving 16 supercharges.
Hence we can study 4d $\mathcal{N}=4$ SYM
by using essentially the same code
as the one used in the works on the 1d SUSY gauge theory \cite{AHNT,%
Hanada:2008gy,Hanada:2008ez,Hanada:2009ne,Hanada:2011fq}.
There are also hybrid formulations of $\mathcal{N}=4$ SYM
preserving two SUSYs,
which use a 2d lattice and fuzzy spheres \cite{Hanada:2010kt,Hanada:2010gs}.
It is claimed that 4d $\mathcal{N}=4$ SYM can be obtained
even at finite $N$ without fine-tuning
if the commutative limit of $\mathcal{N}=4$
\emph{non-commutative} SYM is smooth.
Refs.~\cite{Berenstein:2007wz,Berenstein:2008jn,Berenstein:2010dg}
studied a large-$N$ matrix model of six commuting matrices
as a simplified model for the low-energy dynamics
of 4d $\mathcal{N}=4$ SYM at strong coupling.
It is found \cite{Berenstein:2008jn} that
the three-point functions of half-BPS operators
calculated in
the simplified model
agree with the predictions from
the AdS/CFT correspondence
despite the simplification.

%
%


The aim of this work is to
perform full nonperturbative calculations in
4d $\mathcal{N}=4$ SYM
respecting 16 SUSYs by taking advantage of the large-$N$ limit
based on the proposal in ref.~\cite{Ishii:2008ib}.
In fact it is well known that
the large-$N$ limit
results in
considerable simplification
such as the dominance of planar diagrams and
the factorization property of multi-trace correlation functions.
Under certain assumptions,
one can also show
the equivalence of a theory
to a model with the space-time reduced to a point \cite{EK},
which is commonly referred to as the large-$N$ reduction.
In the reduced model,
the space-time degrees of freedom
are represented
within the internal space,
which is possible since
the internal space becomes infinite dimensional in the large-$N$ limit.
This, in particular,
suggests \cite{Gross}
the possibility of a new regularization scheme
alternative to the lattice that works in the large-$N$ limit.
However, the assumptions for the equivalence
do not hold in many interesting cases.
For instance, the original Eguchi-Kawai model \cite{EK}
for $D$-dimensional SU($N$) pure gauge theory does not work
in $D \ge 3$ due to the spontaneous breaking of
the center symmetry \cite{Bhanot},
which led to various
proposals \cite{Gross,Bhanot,Parisi,Das,GonzalezArroyo,%
Narayanan:2003fc,Kovtun:2007py}.

Another idea for a new type of regularization is the
fuzzy sphere \cite{Madore}.
%
%
%
In this approach, one allows the space-time to have non-commutativity,
and represents the fields by matrices.
Making the matrices finite-dimensional corresponds
to truncating the angular momentum
on the sphere.
This can be done without breaking the gauge symmetry
thanks to the space-time non-commutativity.
Moreover, the SO(3) rotational symmetry, which is the counterpart
of the translational symmetry on a flat space,
is not broken by the regularization.
Therefore, one might expect to obtain a regularization which does not
break SUSY.
In general, it is not possible to
get rid of the space-time non-commutativity
in the continuum limit due to
the so-called UV/IR mixing \cite{Minwalla:1999px}.
However, since the space-time non-commutativity does not affect the
planar diagrams,
one can still think of a new regularization
that works in the large-$N$ limit of the original field theory.

Extending these ideas further,
it was proposed \cite{Ishii:2008ib}
that the ${\cal N}=4$ SU($N$) SYM on $R\times S^3$
can be regularized
in the planar large-$N$ limit preserving 16 SUSYs.
This proposal extends the idea of large-$N$ reduction
to a curved space, while solving the problem of the center
symmetry breaking in the original idea.
The number of preserved SUSYs is
half of the original 32 SUSYs, but
we may consider it optimal in the sense that the conformal symmetry,
which would enhance the symmetry from
16 SUSYs to 32 SUSYs, is inevitably broken
by any regularization.
Following the spirit of the large-$N$ reduction,
let us first collapse the $S^3$ to a point.
The one-dimensional gauge theory obtained in this way \cite{Kim-Klose-Plefka}
is nothing but
the plane wave matrix model (PWMM) \cite{Berenstein:2002jq},
which has 16 SUSYs.
The model possesses many classical vacua,
all of which preserve the 16 SUSYs of the PWMM.
In particular, there exist
%
classical
vacua, which correspond to
multi-fuzzy-sphere configurations
with different radii.
By considering the PWMM around such a classical vacuum,
one can retrieve the theory
before dimensional reduction in the planar
limit \cite{Ishii:2008ib}.\footnote{See
refs.\ \cite{Ishiki:2006rt,Ishiki:2006yr,Ishii:2007ex,Ishii:2008tm}
for earlier discussions that led to this proposal.
This equivalence was
confirmed at finite temperature in the weak coupling
regime \cite{Ishiki:2008te,Ishiki:2009sg,Kitazawa:2008mx}.
It has also been extended to general group manifolds and
coset spaces \cite{Kawai:2009vb,Kawai:2010sf}.
The large-$N$ reduction for Chern-Simons theory
on $S^3$ is demonstrated in refs.~\cite{Ishiki:2009vr,Ishiki:2010pe}.
The large-$N$ reduction for ${\cal N}=4$ SYM has been confirmed
for the circular Wilson loop to all orders of
perturbation theory \cite{Ishiki:2011ct}.
More recently, the localization method \cite{Pestun:2007rz}
has been used to demonstrate the large-$N$ reduction
for ${\cal N}=4$ SYM \cite{Asano:2012zt}
and three-dimensional ${\cal N}=2$ theories \cite{Asano:2012gt,Honda:2012ni}.
}

We use this proposal
to study the ${\cal N}=4$ SU($N$) SYM on $R\times S^3$
on a computer.
Since the theory is conformally invariant,
one can actually map the theory on to $R^4$,
where various predictions are available.
Here we calculate correlation functions
of chiral primary operators (CPOs), and compare the results
against
the predictions from
the AdS/CFT correspondence.\footnote{See
ref.\ \cite{Nishimura:2009xm,Honda:2011qk}
for some preliminary results on
the Wilson loop.}
In particular, we find that the two-point and three-point functions
agree with the corresponding free theory results
even at strong coupling
up to overall constant factors,
which can be absorbed consistently by appropriate normalization
of the operator.
This implies that a weaker version of the SUSY non-renormalization
property holds even at the regularized level.\footnote{The
non-renormalization property
actually refers to a stronger statement that
the agreement with the free theory results
should hold without any constant factors.
This is expected to be realized
only by taking the continuum and infinite-volume limits
in the present approach.}
In contrast, our results for the four-point functions
show clear deviation from
the SUSY non-renormalization property,
which \emph{cannot} be absorbed by the same
normalization of the operator.
%
Indeed this deviation turns out to be consistent
with the prediction from
the AdS/CFT correspondence.

The rest of this paper is organized as follows.
In section \ref{sec:corr-fun-sym}
we review some known results for the correlation functions
of CPOs
in ${\cal N}=4$ SYM on $R^4$
obtained
from gauge theory analyses and from the AdS/CFT correspondence.
In section \ref{sec:large-N-red}
we explain the conformal map between
${\cal N}=4$ SYM on $R^4$ and that on $R\times S^3$,
and review the large-$N$ equivalence of the latter theory
to the PWMM.
In section \ref{sec:corr-fun-pwmm}
we define the correlation functions of CPOs
in ${\cal N}=4$ SYM on $R\times S^3$
and discuss their relationship to the correlation functions
in the PWMM.
We also confirm this relationship
by explicit calculations in the free theory case.
%
In section \ref{sec:monte}
we explain our numerical method for calculating
the correlation functions in the PWMM.
In section \ref{sec:results} we present our results and compare
them against the predictions
from
the AdS/CFT correspondence.
Section \ref{sec:summary} is devoted to a summary and discussions.
In appendix \ref{appendix:A} we present the prediction for
the four-point function obtained from the AdS/CFT correspondence.
In appendix \ref{appendix:B} we explain how we
evaluate the predicted four-point function
in the form that can be compared
with the Monte Carlo data directly.
In appendix \ref{appendix:E} we present some useful formulae
for fuzzy spherical harmonics, which are used to evaluate
the correlation functions in the PWMM in the free theory case.
In appendix \ref{appendix:F} we present free theory results
for the correlation functions in the PWMM with
finite regularization parameters.
In appendix \ref{appendix:C} we discuss the
stability of the background, which is crucial for the large-$N$
reduction to work.
In appendix \ref{appendix:D} we discuss the dependence of
our results on the regularization parameters.

\section{Correlation functions of
CPOs
in ${\cal N}=4$ SYM on $R^4$}
\label{sec:corr-fun-sym}

In this section we review some important properties of
CPOs
and their correlation functions in ${\cal N}=4$ SYM on $R^4$.
We also discuss a prediction for the four-point correlation function
of CPOs obtained from the AdS/CFT correspondence.

The action of ${\cal N}=4$ ${\rm U}(N)$ SYM on $R^4$ is given by
\begin{align}
S=\frac{1}{g_{\rm YM}^2}\int d^4x \,  \mbox{tr}
\left(\frac{1}{4} \, F_{\mu\nu}^2  + \frac{1}{2} \, (D_{\mu}\phi_a)^2
-\frac{1}{4} \, [\phi_a,\phi_b]^2\right) + \ldots
\label{action of N=4 SYM on R^4}
\end{align}
with $D_{\mu}=\partial_{\mu}+i \, [A_{\mu}, \;\;]$
and $F_{\mu\nu}=\partial_{\mu}A_{\nu}-\partial_{\nu}A_{\mu}
+i \, [A_{\mu},A_{\nu}]$,
where we have omitted the fermionic terms.
This theory has ${\rm PSU}(2,2|4)$ superconformal symmetry
with 32 supercharges.
There are six scalars $\phi_a$ ($a=4, \ldots , 9$),
which transform as the fundamental representation under
the R-symmetry ${\rm SO}(6)_{\rm R}$.
CPOs in ${\cal N}=4$ SYM on $R^4$ are the operators of the form
\begin{align}
{\cal O}_{I} (x) =
T_I^{a_1a_2\cdots a_{\Delta_I}}
\mbox{tr} \Bigl(\phi_{a_1}(x) \phi_{a_2}(x) \cdots \phi_{a_{\Delta_I}}(x) \Bigr) \ ,
\label{definition of CPO}
\end{align}
where $T_I^{a_1a_2\cdots a_{\Delta_I}}$ is
a totally symmetric traceless tensor of ${\rm SO}(6)_{\rm R}$
with rank $\Delta_I$.
Since these operators ${\cal O}_{I}$ are half-BPS,
their dimensions are not renormalized and therefore they are equal
to the canonical dimensions $\Delta_I$.
For convenience, we choose $T_I$ to be orthonormal.

We consider the $n$-point correlation
functions of CPOs defined by
\begin{align}
\Bigl\langle {\cal O}_{I}(x_1){\cal O}_{J}(x_2)\cdots
{\cal O}_{K}(x_n) \Bigr\rangle \ ,
\label{n-point function}
\end{align}
where $x_1,x_2,\cdots,x_n \in R^4$.
%
The one-point function of CPO vanishes due to the conservation of
the R-charge.
For the same reason, the two-point and three-point functions
of CPOs have only connected contribution, while
the four-point functions can also have disconnected contribution,
which is factorized into a product of two-point functions.
There are two classes of correlators of the form (\ref{n-point function})
with special properties, which are
the extremal correlators characterized by
$\Delta_{I}=\Delta_{J}+\cdots+\Delta_{K}$
and the next-to-extremal correlators characterized by
$\Delta_{I}=\Delta_{J}+\cdots+\Delta_{K}-2$.
%
In fact a field theoretical analysis strongly suggests
that the non-renormalization property holds for
the extremal correlator \cite{Bianchi:1999ie,Eden:1999kw,Eden:2000gg} and
the next-to-extremal correlator \cite{Erdmenger:1999pz,Eden:2000gg}.

The space-time dependence of
the two-point and three-point functions of CPOs is completely
determined by the conformal symmetry as
\begin{align}
&\Bigl\langle {\cal O}_{I}(x_1){\cal O}_{J}(x_2)\Bigr\rangle
=\frac{C_{I}(g_{\rm YM},N)}{(x_{12})^{2\Delta_I}}
\delta_{IJ} \ ,\nonumber\\
&\Bigl\langle{\cal O}_{I}(x_1){\cal O}_{J}(x_2){\cal O}_{K}(x_3)\Bigr\rangle
=\frac{C_{IJK}(g_{\rm YM},N)}{(x_{12})^{\Delta_I+\Delta_J-\Delta_K}
(x_{23})^{-\Delta_I+\Delta_J+\Delta_K}
(x_{31})^{\Delta_I-\Delta_J+\Delta_K}} \ ,
\end{align}
where $x_{ij}=|x_i-x_j|$.
Let us
define the ratios of the
two-point and three-point correlation functions
to those in the free theory of ${\cal N}=4$ SYM as
\begin{align}
&\frac{\Bigl\langle {\cal O}_{I}(x_1){\cal
      O}_{I}(x_2)
\Bigr\rangle}
{\Bigl\langle {\cal O}_{I}(x_1){\cal O}_{I}(x_2)\Bigr\rangle_{\rm free}}
=\frac{C_{I}(g_{\rm YM},N)}{C_{I}(g_{\rm YM},N)_{\rm free}}
\equiv c_{I}(g_{\rm YM},N) \ ,
\label{constant2pt} \\
&\frac{\Bigl\langle {\cal O}_{I}(x_1)
{\cal O}_{J}(x_2) {\cal O}_{K}(x_3)\Bigr\rangle}
{\Bigl\langle {\cal O}_{I}(x_1){\cal O}_{J}(x_2){\cal O}_{K}(x_3)
\Bigr\rangle_{\rm free}}
=\frac{C_{IJK}(g_{\rm YM},N)}{C_{IJK}(g_{\rm YM},N)_{\rm free}}
\equiv c_{IJK}(g_{\rm YM},N) \ ,
\label{constant3pt}
\end{align}
where
the suffix ``free'' represents a quantity defined for the free theory.

There is a strong evidence from a field theoretical analysis
that the non-renormalization property holds for the two-point and three-point
functions of
CPOs \cite{D'Hoker:1998tz,Howe:1998zi,Skiba:1999im,GonzalezRey:1999ih,%
Bianchi:2000hn,Penati:1999fr,%
Eden:1999gh,Arutyunov:2001qw,Heslop:2001gp}.\footnote{Since
non-vanishing two-point correlation functions are always extremal,
the argument in ref.~\cite{Bianchi:1999ie,Eden:1999kw,Eden:2000gg}
implies their non-renormalization.}
Namely,
\begin{align}
c_{I}(g_{\rm YM},N)=1 \ , \quad
c_{IJK}(g_{\rm YM},N)=1 \ .
\label{non-renormalization of 3-pt fn}
\end{align}
Moreover, one can make an independent
argument for the non-renormalization
of the two-point and three-point correlation functions
of the CPOs with $\Delta=2$,
using the fact that these CPOs belong to the same multiplet
as the stress tensor and
the R-symmetry current
\cite{Gubser:1997se,Anselmi:1997am,%
Freedman:1998tz,Lee:1998bxa}.\footnote{Since the three-point correlation
function of these CPOs is next-to-extremal,
the argument in ref.~\cite{Erdmenger:1999pz,Eden:2000gg}
implies its non-renormalization.}
On the other hand, the prediction from the AdS/CFT correspondence
for general two-point and three-point
correlation functions in the planar limit at strong coupling
is given through the GKP-Witten relation \cite{Gubser:1998bc,Witten:1998qj}
as
\begin{align}
\left.\frac{c_{IJK}}
{\sqrt{c_{I}c_{J}c_{K}}}\right|_{N\rightarrow\infty,\lambda\rightarrow\infty}
=1 \ ,
\label{gravity prediction}
\end{align}
which is consistent with
(\ref{non-renormalization of 3-pt fn}).
Here the $N\rightarrow \infty$ limit is taken with
fixed 't Hooft coupling
$\lambda \equiv g_{\rm YM}^2 N$, which corresponds to the
planar limit, and the $\lambda \rightarrow \infty$ limit is taken afterwards.

The space-time dependence of the four-point correlation functions
is not completely determined by the symmetry, but they can be written as
\begin{align}
& \Bigl\langle {\cal O}_{I}(x_1){\cal O}_{J}(x_2) {\cal O}_{K}(x_3)
{\cal O}_{L}(x_4)
\Bigr\rangle
=\frac{C_{IJKL}(g_{\rm YM},N;u,v)}
{(x_{13})^{\Delta_I+\Delta_K} (x_{24})^{\Delta_J+\Delta_L}} \ ,
\end{align}
where $u$ and $v$ are
two independent conformal invariants defined by
\begin{align}
&u = \frac{(x_{12})^2 (x_{34})^2}{(x_{13})^2 (x_{24})^2} \ , \quad \quad
v = \frac{(x_{12})^2 (x_{34})^2}{(x_{14})^2 (x_{23})^2} \ .
\label{cross-ratios}
\end{align}
Note, in particular, that the four-point correlation function
of CPOs with $\Delta=2$
is neither extremal nor next-to-extremal,
and therefore it is expected to be renormalized.

\begin{figure}[t]
\begin{center}
\includegraphics[width=10cm]{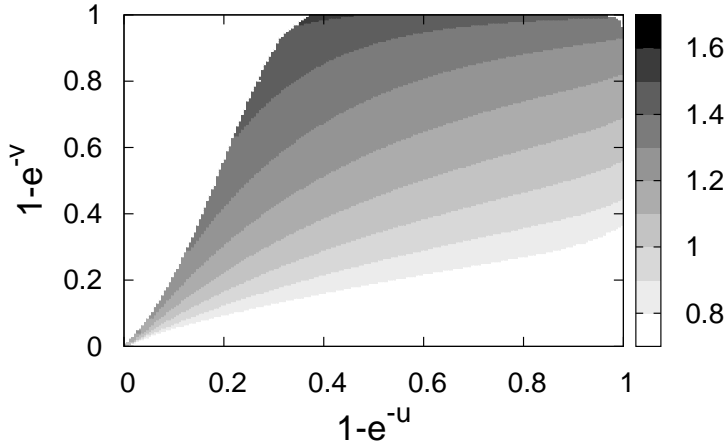}
\end{center}
\caption{
The contour lines of the function $c(u,v)$
in (\ref{ads-cft-pred-4pt})
predicted by the AdS/CFT correspondence
are plotted against
$\bar{u}=1-e^{-u}$ and $\bar{v}=1-e^{-v}$.
The range of the function is $ 0.80 \lesssim c(u,v) \lesssim 1.66 $.
The blank regions near $(\bar{u},\bar{v})=(1,0),(0,1),(1,1)$ represent
the region of $(u,v)$ that cannot be realized
according to the definition (\ref{cross-ratios}).
See footnote 6.
%
}
\label{fourpt-contourplot}
\end{figure}

In this paper we focus on the CPOs
with the lowest dimension $\Delta=2$ given by
\begin{align}
{\cal O}_{ab}(x)
=\mbox{tr} \Bigl(\phi_a(x) \phi_b(x) \Bigr) \;\;\mbox{with}\;\; a\neq b \ .
\label{chiral primary operator with lowest dimension on R^4}
\end{align}
According to the above general argument,
the correlation functions of the CPOs
(\ref{chiral primary operator with lowest dimension on R^4})
can be parametrized as
\begin{align}
\Bigl\langle {\cal O}_{45}(x_1){\cal O}_{54}(x_2) \Bigr\rangle
&= \frac{\lambda^2}{(4 \pi ^2 )^2}
\frac{c^{(2)}(\lambda , N )}{(x_{12})^4} \ ,
\label{CPO-cor-def-2pt} \\
\Bigl\langle{\cal O}_{45}(x_1){\cal O}_{56}(x_2){\cal O}_{64}(x_3)\Bigr\rangle
&=
\frac{\lambda^3}{(4 \pi ^2 )^3 N }
\frac{c^{(3)}(\lambda , N )}{(x_{12})^2(x_{23})^2(x_{31})^2} \ ,
\label{CPO-cor-def-3pt} \\
\Bigl\langle {\cal O}_{45}(x_1)
{\cal O}_{56}(x_2){\cal O}_{67}(x_3) {\cal O}_{74}(x_4)\Bigr\rangle
&=
\frac{\lambda^4}{(4 \pi ^2 )^4 N^2 }
\frac{c^{(4)}(\lambda, N;u,v)}{(x_{12})^2 (x_{23})^2 (x_{34})^2 (x_{41})^2} \ .
\label{4-point function}
\end{align}
These correlation functions have only connected contribution.
By using the propagators for the scalar fields
\begin{align}
\Bigl\langle \phi_a(x)_{pq}\phi_b(y)_{rs} \Bigr\rangle
=\frac{g_{\rm YM}^2\delta_{ab}\delta_{ps}\delta_{qr}}{4\pi^2|x-y|^2} \ ,
\label{propagator on R^4}
\end{align}
where $p,q,r,s=1,\cdots, N$, one finds that the
coefficients are given for the free theory by
\begin{align}
\lim_{\lambda \rightarrow 0 } c^{(2)}(\lambda , N )
= \lim_{\lambda \rightarrow 0 }  c^{(3)}(\lambda , N )
= \lim_{\lambda \rightarrow 0 } c^{(4)}(\lambda, N;u,v) = 1  \ .
\label{free-coeff-c}
\end{align}

On the other hand, the AdS/CFT correspondence predicts that
\begin{align}
\label{def-c2}
& \lim_{N \rightarrow \infty ,\lambda \rightarrow \infty }
  c^{(2)}(\lambda , N )
= \zeta   \ ,  \\
\label{def-c3}
& \lim_{N \rightarrow \infty ,\lambda \rightarrow \infty }
  c^{(3)}(\lambda , N )
= \zeta^{3/2}  \ ,  \\
& \lim_{N \rightarrow \infty ,\lambda \rightarrow \infty }
  c^{(4)}(\lambda, N;u,v)
= \zeta^{2} c(u,v)  \ .
\label{ads-cft-pred-4pt}
\end{align}
Here $\zeta$ is a parameter associated with the normalization of the
operator, which cannot be fixed by the AdS/CFT correspondence,
while the non-renormalization property (\ref{non-renormalization of 3-pt fn})
implies $\zeta =1$.
The prediction for $c(u,v)$ in (\ref{ads-cft-pred-4pt})
is obtained in ref.~\cite{Arutyunov:2000py},
and it is shown in
figure~\ref{fourpt-contourplot}.\footnote{Note that
the two parameters $u$ and $v$ defined by eq.~(\ref{cross-ratios}) cannot
take arbitrary set of values within $0 \le u, v \le \infty$.
For instance, since $v \rightarrow 0$ as $u \rightarrow 0$,
we cannot take the $u \rightarrow 0$ limit while keeping $v$ finite,
and vice versa.
Similarly, one cannot take the $v \rightarrow \infty$ limit
for $u \neq 1$ and vice versa.
\label{foot:c-bound}}
One finds that the function
$c(u,v)$ has nontrivial dependence on $u$ and $v$,
and in particular, it differs from 1, which implies
that the four-point function is renormalized.
This is consistent with the fact that
the four-point function considered here
is neither extremal nor next-to-extremal.
%
The explicit form of $c(u,v)$
and useful expressions for its evaluation are given
in appendix \ref{appendix:A}.

\section{${\cal N}=4$ SYM on $R\times S^3$ and its large-$N$ reduction}
\label{sec:large-N-red}
In this section we review a non-perturbative formulation of
planar ${\cal N}=4$ SYM on $R\times S^3$ proposed
in ref.~\cite{Ishii:2008ib}.
(See
ref.~\cite{Ishiki:2011ct} for a review.)
This formulation is based on an extension of the large-$N$ reduction \cite{EK}
to a curved manifold $S^3$.
In section \ref{sec:SYMr4rxs3}
we review the conformal map between
${\cal N}=4$ SYM on $R^4$ and that on $R\times S^3$.
In section \ref{sec:pwmm-from-rxs3}
we show how the PWMM is obtained from
${\cal N}=4$ SYM on $R\times S^3$ through a dimensional reduction.
In section \ref{sec:SYMrxs3-from-pwmm}
we discuss how ${\cal N}=4$ SYM on $R\times S^3$ is
retrieved from the PWMM through the novel large-$N$ reduction.


\subsection{Conformal map between
${\cal N}=4$ SYM on $R^4$ and that on $R\times S^3$}
\label{sec:SYMr4rxs3}

It is well known that ${\cal N}=4$ SYM on $R^4$
has a moduli space, which is characterized by
the vacuum expectation values of the scalar fields.
The conformal symmetry is spontaneously broken
except at the conformal point, which corresponds to a point in
the moduli space where the vacuum expectation values all vanish.
In section \ref{sec:corr-fun-sym} we implicitly assumed that the theory
is defined at the conformal point.
In fact ${\cal N}=4$ SYM on $R^4$ at the conformal point
is equivalent to ${\cal N}=4$ SYM on $R\times S^3$ through
the conformal map.
To see that, we apply a Weyl transformation defined by
\begin{align}
A_{\mu}  \mapsto A_{\mu} \  , \quad
\phi_a \mapsto e^{-\frac{\rho(x)}{2}}\phi_a \ , \quad
\delta_{\mu\nu} \mapsto g_{\mu\nu}=e^{\rho(x)}\delta_{\mu\nu}
\label{Weyl transformation}
\end{align}
to the action (\ref{action of N=4 SYM on R^4})
of ${\cal N}=4$ SYM on $R^4$.
This gives rise to ${\cal N}=4$ SYM on a curved space
endowed with a metric $g_{\mu\nu}$ with the action
\begin{align}
S=\frac{1}{g_{\rm YM}^2}\int d^4x \sqrt{g} \, \mbox{tr}
\left(\frac{1}{4}g^{\mu\lambda}g^{\nu\rho}F_{\mu\nu}F_{\lambda\rho}
+\frac{1}{2}g^{\mu\nu}D_{\mu}\phi_aD_{\nu}\phi_a
+\frac{1}{12}{\cal R} \phi_a^2-\frac{1}{4}[\phi_a,\phi_b]^2\right) \ ,
\label{action of N=4 SYM on curved space}
\end{align}
where ${\cal R}$ is the Ricci scalar constructed from $g_{\mu\nu}$.

Rewriting
the metric on $R^4$ in the polar coordinates as
\begin{align}
ds_{R^4}^2=dx^{\mu}dx^{\mu}
          =dr^2 +r^2 d\Omega_3^2
          =e^{\mu t}\left(dt^2
            +\left(\frac{2}{\mu}\right)^2 d\Omega_3^2 \right)
          =e^{\mu t}ds^2_{R\times S^3} \  ,
\label{metric}
\end{align}
where $r=\frac{2}{\mu}e^{\frac{\mu}{2}t}$,
one finds that
$R^4$ is transformed to $R\times S^3$ by a Weyl transformation
with $\rho(x)$ in (\ref{Weyl transformation})
given by $\rho=-\mu t$.
The radius of the resulting $S^3$ is given by
\begin{align}
R_{S^3}=\frac{2}{\mu} \  .
\label{R-mu-rel}
\end{align}
Thus we have seen that
${\cal N}=4$ SYM on $R^4$ at the conformal point is equivalent to
${\cal N}=4$ SYM on $R\times S^3$.

The relation between the Cartesian coordinates
and the polar coordinates of $R^4$ in (\ref{metric})
is given, for instance, as
\begin{align}
x^1 &=r \, \cos\frac{\theta}{2} \, \cos\frac{\varphi+\psi}{2} \ , \quad
x^2 =r \, \cos\frac{\theta}{2} \, \sin\frac{\varphi+\psi}{2} \ , \nonumber\\
x^3 & =r \, \sin\frac{\theta}{2} \, \cos\frac{-\varphi+\psi}{2} \ , \quad
x^4 =r \, \sin\frac{\theta}{2} \, \sin\frac{-\varphi+\psi}{2} \ ,
\label{R^4 and RxS^3}
\end{align}
where $0\leq\theta\leq\pi$, $0\leq\varphi<2\pi$ and $0\leq\psi<4\pi$,
which we use later.

\subsection{PWMM from ${\cal N}=4$ SYM on $R\times S^3$}
\label{sec:pwmm-from-rxs3}

Let us write down the action
(\ref{action of N=4 SYM on curved space})
explicitly in the case of $R\times S^3$.
For that purpose we recall that $S^3$ can be viewed as
the ${\rm SU}(2)$ group manifold. The isometry of $S^3$ corresponds to
the left and right translations on the ${\rm SU}(2)$ group manifold
as one can see from the relation
${\rm SO}(4)={\rm SU}(2)\times {\rm SU}(2)$.
As is well known,
one can construct the right-invariant 1-forms
$e^i_{\bar{\mu}}$,
where $i=1,2,3$ and $\bar{\mu}$ stands for $\theta,\phi,\psi$.
(Note that $\mu$ in (\ref{action of N=4 SYM on curved space})
represents either $t$ or $\bar{\mu}$.)
The 1-forms $e^i_{\bar{\mu}}$ are called
``dreibein'' and satisfy the Maurer-Cartan equation
\begin{align}
\partial_{\bar{\mu}} \, e^i_{\bar{\nu}}
- \partial_{\bar{\nu}} \, e^i_{\bar{\mu}}
-\mu \, \epsilon_{ijk} \, e^j_{\bar{\mu}} \, e^k_{\bar{\nu}}=0 \ .
\label{Maurer-Cartan equation}
\end{align}
By using the inverse of $e^i_{\bar{\mu}}$ denoted by
$e_i^{\bar{\mu}}$, one can also construct the Killing
vector\footnote{For details such as
an explicit form of $e^i_{\bar{\mu}}$ and
${\cal L}_i$, see refs.~\cite{Ishii:2008ib,Ishii:2008tm}.}
\begin{align}
{\cal L}_i=-\frac{i}{\mu}
\, e^{\bar{\mu}}_i \, \partial_{\bar{\mu}} \ ,
\label{Killing vector}
\end{align}
which represents the generators of the left translation and satisfies
the ${\rm SU}(2)$ algebra
$[{\cal L}_i,{\cal L}_j]=i\epsilon_{ijk}{\cal L}_k$.
The ${\rm SO}(4)$-isometric metric on $S^3$ is given in terms of
$e^i_{\bar{\mu}}$ as
\begin{align}
g_{\bar{\mu}\bar{\nu}}=
e^i_{\bar{\mu}} \, e^i_{\bar{\nu}} \ ,
\label{metric of S^3}
\end{align}
and the Ricci scalar constructed from $g_{\bar{\mu}\bar{\nu}}$ takes
a constant value $\frac{3}{2}\, \mu^2$.

We expand the gauge field on $S^3$
with respect to
$e^i_{\bar{\mu}}$ as
$A_{\bar{\mu}}=e^i_{\bar{\mu}}X_i$,
and denote $\phi_a$ as $X_a$.
Then, using (\ref{Maurer-Cartan equation}), (\ref{Killing vector}) and
(\ref{metric of S^3}),
we can rewrite
(\ref{action of N=4 SYM on curved space}) in terms of $X_i$
and $X_a$ as
\begin{align}
S &=\frac{V_{S^3}}{g_{\rm YM}^2}\int dt \, \frac{d\Omega_3}{2\pi^2}
\, \mbox{tr} \left[ \frac{1}{2}\, (D_{t}X_i-i\mu{\cal L}_iA_t)^2
+\frac{1}{2}\, \left\{\mu X_i+i\epsilon_{ijk}
\left(\mu{\cal L}_jX_k+\frac{1}{2}[X_j,X_k]\right)\right\}^2 \right.
\nonumber\\
&\qquad\qquad\qquad\qquad\;\;\;\left.
+\frac{1}{2}\, (D_tX_i)^2
-\frac{1}{2}\, (\mu{\cal L}_iX_a+[X_i,X_a])^2
+\frac{1}{8}\, \mu^2 X_a^2
-\frac{1}{4} \, [X_a,X_b]^2 \right] \ ,
\label{action of N=4 SYM on RxS^3}
\end{align}
where $d\Omega_3=\frac{1}{8}\sin\theta \, d\theta \, d\phi \, d\psi$ is
the volume element of a unit $S^3$
and the covariant derivative is defined by
$D_t=\partial_t + i \, [A_t, \ \cdot \ ]$.

Let us perform a dimensional reduction over
$S^3$ \cite{Kim-Klose-Plefka,Lin:2005nh}
by making the fields
depend only on $t$, namely by
setting ${\cal L}_iA_t=0$, ${\cal L}_iX_i=0$
and ${\cal L}_iX_a=0$.
The resulting theory is nothing but the PWMM,
and its complete form including the fermion fields is
\begin{align}
S_{\rm PW}
= \frac{1}{g_{\rm PW}^2}
\int
dt \, \mbox{tr} &
\left(\frac{1}{2}\, (D_{t}X_M)^2
-\frac{1}{4} \, [X_M,X_N]^2
+\frac{1}{2} \, \mu^2 X_i^2
+\frac{1}{8} \, \mu^2 X_a^2
+ i \, \mu \, \epsilon_{ijk} \, X_iX_jX_k \right. \nonumber\\
&\left.+\frac{1}{2} \, \Psi^{\dagger} D_t \Psi
-\frac{1}{2} \, \Psi^{\dagger}\gamma_M[X_M,\Psi]
+\frac{3}{8} \, i  \,  \mu \, \Psi^{\dagger}\gamma_{123}\Psi
\right) \ ,
\label{pp-action}
\end{align}
where $M$ runs from 1 to 9.
This is a consistent truncation in the sense that every classical
solution in (\ref{pp-action}) is a classical solution
in (\ref{action of N=4 SYM on RxS^3}).
The PWMM has the ${\rm SU}(2|4)$ symmetry with 16 supercharges, which is
a subgroup of ${\rm PSU}(2,2|4)$ of ${\cal N}=4$ SYM.
By setting $\mu=0$ in eq.~(\ref{pp-action}),
one obtains the Matrix Theory \cite{BFSS}.
In fact the PWMM can be viewed
as a mass deformation of the Matrix Theory
preserving its SUSYs completely.

The PWMM has many discrete vacua
given by
\begin{align}
X_i  =  \mu L_i =\mu \bigoplus_{I=1}^{\nu}
\Bigl( L_i^{(n_I)}\otimes {\bf 1}_{k_I} \Bigr) \ , \quad \quad
X_a = 0 \ ,
 \label{background}
\end{align}
where $L_i^{(r)}$ are
the $r$-dimensional irreducible representation of
the SU$(2)$ generators obeying
\begin{align}
[L_i^{(n)},L_j^{(n)}]=i \, \epsilon_{ijk} \, L_k^{(n)} \ .
\end{align}
The parameters $n_I$ and $k_I$ in (\ref{background})
have to satisfy the relation
\begin{align}
\sum_{I=1}^{\nu}n_Ik_I=N \ .
\end{align}
These vacua preserve the SU$(2|4)$ symmetry,
and they are all degenerate.
Since the generators $L_i^{(n)}$ satisfy
$\sum_{i=1}^3 (L_i^{(n)})^2 = \frac{1}{4} (n^2-1) \,  {\bf 1}_n$,
they represent a fuzzy sphere \cite{Madore}.
The classical vacua (\ref{background}) therefore
represent multi-fuzzy-sphere configurations with different radii
$\frac{\mu}{2} \sqrt{(n_I)^2-1}$ and multiplicity $k_I$.

\subsection{${\cal N}=4$ SYM on $R\times S^3$ from PWMM}
\label{sec:SYMrxs3-from-pwmm}

In order to retrieve ${\cal N}=4$ SYM on $R \times S^3$
from its reduced model (\ref{pp-action})
in the planar limit,
we need to pick up a particular background
from (\ref{background})
and consider the theory (\ref{pp-action}) around it \cite{Ishii:2008ib}.
Let us consider the background (\ref{background}) with
\begin{align}
k_I=k \ , \quad n_I=n+I-\frac{\nu+1}{2} \quad \quad
\mbox{for $I =1, \cdots , \nu$}
\label{our background}
\end{align}
and take the large-$N$ limit in such a way that
\begin{align}
k\rightarrow\infty, \;\;
\frac{n}{\nu}\rightarrow\infty, \;\;\nu\rightarrow \infty,\;\;
\mbox{with} \;\;
\tilde{\lambda}
\equiv \frac{g_{\rm PW}^2 k}{n}
\; \; \mbox{fixed} \ .
\label{limit}
\end{align}
Note that the $k\rightarrow\infty$ limit
ensures the dominance of planar diagrams in the reduced model.
Then the resulting theory is equivalent to the planar limit
of ${\cal N}=4$ SYM on $R\times S^3$
with the radius (\ref{R-mu-rel}) of $S^3$
and the 't Hooft coupling constant given
by
\begin{align}
 \lambda
=
\tilde{\lambda}
 \, V_{S^3} \ ,
\label{'t Hooft coupling}
\end{align}
where $V_{S^3}= 2\pi^2 (R_{S^3})^3 = 16 \pi^2 /\mu^3$
is the volume of $S^3$.
%

This equivalence may be viewed as an extension of the large-$N$ reduction
to a curved space $S^3$.
The original idea \cite{EK}
for theories compactified on a torus can fail due to
the instability of the ${\rm U}(1)^D$ symmetric
vacuum of the reduced model \cite{Bhanot}.
This problem is avoided in the above new proposal
since the PWMM is a massive theory
and the vacuum preserves the maximal SUSY.
The instanton transition to other vacua is suppressed
due to the $k\rightarrow \infty$
limit.\footnote{The transition amplitude to other vacua behaves as
$\sim \exp(-{\rm const.} \frac{k^2\nu}{\lambda})$ \cite{Lin:2006tr}.}
Note also that the ``fuzziness" of the spheres disappears
in the $k\rightarrow \infty$ limit
since the UV/IR mixing effects come only
from non-planar diagrams.
The SU($N$) gauge symmetry
of the PWMM (with the gauge function depending only on $t$)
is translated into the four-dimensional SU($k$)
gauge symmetry of ${\cal N}=4$ SYM.
Viewed as a regularization of ${\cal N}=4$ SYM on $R\times S^3$,
the present formulation
respects the ${\rm SU}(2|4)$ symmetry
with 16 supercharges of the PWMM, and in the limit (\ref{limit})
the symmetry is expected to enhance to the full superconformal
symmetry ${\rm PSU}(2,2|4)$ with 32 supercharges of ${\cal N}=4$ SYM.
Since any kind of UV regularization breaks the conformal symmetry,
this regularization
should be considered
optimal from the viewpoint of preserving SUSY.

\section{Correlation functions of CPOs from PWMM}
\label{sec:corr-fun-pwmm}

In this section we show how one can obtain
the correlation functions of CPOs in ${\cal N}=4$ SYM
by calculating their counterparts
in the PWMM.
In particular, we perform explicit calculations
in the free theory case
and confirm that the results obtained in the limit (\ref{limit})
from the PWMM around the background (\ref{our background})
reproduce the correlation functions of CPOs in ${\cal N}=4$ SYM.

\subsection{Correlation functions of CPOs
in ${\cal N}=4$ SYM on $R\times S^3$}
\label{sec:cpo-rs3}

Let us first define correlation functions of CPOs
in ${\cal N}=4$ SYM on $R\times S^3$.
We find from (\ref{Weyl transformation})
that the six scalar fields on $R^4$ are related
to those on $R\times S^3$ as
\begin{align}
\phi_a(x)=e^{-\frac{\mu}{2}t}\phi_a(t,\Omega_3) \ .
\label{scalar_trans}
\end{align}
Correspondingly, the free propagator (\ref{propagator on R^4})
on $R^4$
is translated to that on $R\times S^3$ as
\begin{align}
\Bigl\langle \phi_a(t,\Omega_3)_{pq} \,
\phi_b(t',\Omega_3')_{rs}\Bigr\rangle
=\frac{g_{\rm YM}^2\delta_{ab}\delta_{ps}\delta_{qr}
  e^{\frac{\mu}{2}(t+t')}} {4\pi^2|x-x'|^2} \ .
\label{propagator on RxS^3}
\end{align}
As in (\ref{definition of CPO}),
CPOs in ${\cal N}=4$ SYM on $R\times S^3$ are defined by
\begin{align}
{\cal O}_I(t,\Omega_3)=T_I^{a_1a_2\cdots a_{\Delta_I}}
\mbox{tr} \Bigl(\phi_{a_1}(t,\Omega_3)
\phi_{a_2}(t,\Omega_3) \cdots \phi_{a_{\Delta_I}}(t,\Omega_3) \Bigr) \ .
\end{align}
In particular, the CPOs on $R\times S^3$ with
the lowest energy\footnote{Note that the dimension on $R^4$
corresponds to the energy on $R\times S^3$.}
corresponding to
(\ref{chiral primary operator with lowest dimension on R^4})
are
\begin{align}
{\cal O}_{ab}(t,\Omega_3)
=\mbox{tr} \Bigl(\phi_a(t,\Omega_3)\phi_b(t,\Omega_3) \Bigr)
\;\;\; \mbox{with} \;\;a\neq b \ .
\label{integrated operator with lowest dimension}
\end{align}
By using (\ref{propagator on RxS^3}),
one obtains, for instance,
the two-point function of ${\cal O}_{ab}(t,\Omega_3)$
in the free theory as
\begin{align}
\Bigl\langle {\cal O}_{ab}(t,\Omega_3) \,
{\cal O}_{ab}(t',\Omega_3') \Bigr\rangle_{\rm free}
=\frac{\lambda^2e^{\mu(t+t')}}{16\pi^4|x-x'|^4} \ .
\label{two point function on RxS^3}
\end{align}

In order to relate these operators to their counterparts
in the PWMM, we need
to integrate
${\cal O}_I(t,\Omega_3)$
over a unit $S^3$ and define an operator
\begin{align}
\bar{{\cal O}}_I (t)
\equiv \int \frac{d\Omega_3}{2\pi^2}
 \, {\cal O}_I (t,\Omega_3) \ , \quad
\label{integrated operator}
\end{align}
and, in particular,
\begin{align}
\bar{{\cal O}}_{ab}(t) \equiv
\int \frac{d\Omega_3}{2\pi^2}
 \, {\cal O}_{ab} (t,\Omega_3)  \ .
\label{integrated operator-ab}
\end{align}
The two-point function of $\bar{{\cal O}}_{ab}(t)$
can be calculated
in the
free theory by using (\ref{R^4 and RxS^3})
and (\ref{two point function on RxS^3}) as
\begin{align}
\Bigl\langle \bar{{\cal O}}_{ab}(t)
\bar{{\cal O}}_{ab}(t') \Bigr\rangle_{\rm free}
&= \frac{\lambda^2}{16\pi^4}e^{\mu(t+t')}
\int \frac{d\Omega_3}{2\pi^2}\frac{1}{|x-x'|^4} \nonumber\\
&=\frac{\lambda^2}{16^2\pi^6}e^{\mu(t+t')}
\int_0^{\pi} \!\!\! d\theta
\int_0^{2\pi} \!\!\! d\varphi
\int_0^{4\pi}  \!\!\!  d\psi \sin\theta
\frac{1}{(r^2+r'{}^2-2rr'\cos\frac{\theta}{2}\cos\frac{\varphi+\psi}{2})^2}
\nonumber\\
&=\frac{\lambda^2\mu^4}{(16\pi^2)^2}\frac{e^{-\mu|t-t'|}}{1-e^{-\mu|t-t'|}} \ ,
\label{two-point function}
\end{align}
where we have fixed $x'$ to $(r',0,0,0)$ without loss of generality.

In a similar manner, the three-point and four-point functions
are calculated in the
free theory as
\begin{align}
&N \Bigl\langle \bar{{\cal O}}_{45}(t_1)
\bar{{\cal O}}_{56}(t_2)
\bar{{\cal O}}_{64}(t_3)\Bigr\rangle_{\rm free}
=-\frac{\lambda^3\mu^6}{(16\pi^2)^3}
\log(1-e^{-\frac{\mu}{2}(|t_1-t_2|+|t_2-t_3|+|t_3-t_1|)}) \ , \nonumber\\
&N^2\Bigl \langle \bar{{\cal O}}_{45}(t_1)\bar{{\cal O}}_{56}(t_2)
\bar{{\cal O}}_{67}(t_3)\bar{{\cal O}}_{74}(t_4)\Bigr\rangle_{\rm free}
=\frac{\lambda^4\mu^8}{(16\pi^2)^4}
\mbox{Li}_2(e^{-\frac{\mu}{2}(|t_1-t_2|+|t_2-t_3|+|t_3-t_4|+|t_4-t_1|)}) \ ,
\label{four-point function}
\end{align}
where $\mbox{Li}_2$ is the dilogarithmic function defined by
$\mbox{Li}_2(z)=-\int_0^z\frac{\log(1-u)}{u}du$.
In fact the
correlation functions
in (\ref{two-point function})
and (\ref{four-point function})
have only planar contribution in the free theory.

\subsection{Corresponding correlation functions in PWMM}
\label{sec:corresponding-PWMM}

As we reviewed in section \ref{sec:SYMrxs3-from-pwmm},
${\cal N}=4$ SYM on $R\times S^3$
is equivalent to the PWMM in the large-$N$ limit.
In this equivalence,
the operators in the PWMM
corresponding to
(\ref{integrated operator}) in ${\cal N}=4$ SYM
are
\begin{align}
{\cal O}_I(t)=T_I^{a_1a_2\cdots a_{\Delta_I}}
\mbox{tr}\left(X_{a_1}(t)X_{a_2}(t)\cdots X_{a_{\Delta_I}}(t)\right)  \ .
\label{CPO in PWMM}
\end{align}
The relationship between the correlation functions is
given by
\begin{align}
N^{M-2} \Bigl\langle \bar{{\cal O}}_{I_1}(t_1)
\cdots\bar{{\cal O}}_{I_M}(t_M) \Bigr\rangle
=\frac{k^{M-2}}{n^M\nu}
\Bigl\langle{\cal O}_{I_1}(t_1)\cdots{\cal O}_{I_M}(t_M)
\Bigr\rangle_{\rm PW} \ ,
\label{correspondence}
\end{align}
where the symbol $\langle  \  \cdot \ \rangle_{\rm PW}$
on the right-hand side represents a VEV with respect to the PWMM
(\ref{pp-action})
around the background (\ref{our background}),
and it is assumed that
the limit (\ref{limit}) is taken.
The trace over $N\times N$ matrices in (\ref{CPO in PWMM}),
which ensures the gauge invariance,
actually corresponds to integrating over $S^3$
and taking the trace over $k\times k$ indices.
Both sides of (\ref{correspondence})
represent planar and connected contribution,
and the factors $N^{M-2}$, $k^{M-2}$ on each side
make the quantities finite in the planar limit.
This kind of correspondence holds for general gauge-invariant
operators in ${\cal N}=4$ SYM on $R\times S^3$.

Here we show explicitly in the free theory
that (\ref{correspondence}) holds
for correlation functions defined
by (\ref{two-point function}) and
(\ref{four-point function}).
Note that the corresponding correlation functions in the PWMM
have only planar contribution in the free theory
similarly to the situation in ${\cal N}=4$ SYM.
In order to calculate the right-hand side of
(\ref{correspondence}) around the
background (\ref{our background})
%
in the PWMM,
we expand the $(I,J)$ block of $X_a$ ---denoted by $X^{(I,J)}_a$---
in terms of the fuzzy spherical harmonics
defined in (\ref{definition of fuzzy spherical harmonics}) as
\begin{align}
X^{(I,J)}_a(t)=\sum_{j=\frac{1}{2}|n_I-n_J|}^{\frac{1}{2}(n_I+n_J)-1}
\sum_{m=-j}^j x^{(I,J)}_{ajm}(t)\otimes Y_{jm}^{(n_I,n_J)} \ .
\end{align}
Note that $X^{(I,J)}_a(t)$ is a $(n_Ik)\times(n_Jk)$ matrix,
while $x^{(I,J)}_{ajm}(t)$ is a $k\times k$ matrix.
Using (\ref{property of fuzzy spherical harmonics}), we find that
the operator corresponding to (\ref{integrated operator-ab})
can be
expressed
as
\begin{align}
{\cal O}_{ab}(t)\equiv \mbox{tr}(X_a(t)X_b(t))
=\sum_{I,J}\sum_{j,m}(-1)^{m-\frac{1}{2}(n_I-n_J)}\mbox{tr}
\Bigl(x_{ajm}^{(n_I,n_J)}x_{bj-m}^{(n_J,n_I)} \Bigr) \ .
\label{operator in PWMM}
\end{align}
Expanding the action (\ref{pp-action})
around the background (\ref{our background}),
we find that the quadratic terms in $X_a$
is diagonalized in terms of
$x_{ajm}(t)$ as
\begin{align}
&\frac{1}{2g_{\rm PW}^2}\int dt \,
\mbox{tr}\left(-X_a\partial_t^2X_a+\mu^2X_a[L_i,[L_i,X_a]]+\frac{\mu^2}{4}X_a^2\right)
\nonumber\\
&=\frac{1}{2g_{\rm PW}^2}\int dt \sum_{I,J}\sum_{j,m}(-1)^{m-\frac{1}{2}(n_I-n_J)}\mbox{tr}\left(x^{(I,J)}_{ajm}
\left(-\partial_t^2+\mu^2\left(j+\frac{1}{2}\right)^2\right)x^{(J,I)}_{aj\:-m}\right),
\label{free part of X_a}
\end{align}
where $L_i$ is given by (\ref{background}).
From (\ref{free part of X_a}),
we can read off the free propagator for $x^{(I,J)}_{ajm}$ as
\begin{align}
\Bigl\langle
\Bigl(  x^{(I,J)}_{ajm}(t)  \Bigr)  _{pq}
\Bigl(x^{(I',J')}_{a'j'm'}(t')  \Bigr)  _{rs} \Bigr\rangle
&  =
\frac{g_{\rm PW}^2(-1)^{m-\frac{1}{2}(n_I-n_J)}}{2\mu(j+\frac{1}{2})}
e^{-\mu(j+\frac{1}{2})|t-t'|}
\nonumber
\\
& \quad \times
\delta_{aa'}\delta_{IJ'}\delta_{JI'}\delta_{jj'}\delta_{m\:-m'}
\delta_{ps}\delta_{qr} \ ,
\label{propagator}
\end{align}
where $0\leq p,q,r,s \leq k$.

Let us calculate the two-point function
corresponding to (\ref{two-point function})
in the free theory.
By using (\ref{propagator}),
we obtain
\begin{align}
\frac{1}{n^2\nu}
\Bigl\langle {\cal O}_{ab}(t){\cal O}_{ab}(t')\Bigr\rangle_{\rm PW,free}
=\frac{g_{\rm PW}^4k^2}{2\mu^2n^2\nu}\sum_{I,J=1}^{\nu}\sum_{j=\frac{1}{2}|n_I-n_J|}^{\frac{1}{2}(n_I+n_J)-1}
\frac{e^{-\mu(2j+1)|t-t'|}}{j+\frac{1}{2}} \ .
\label{lowest order}
\end{align}
We can show that (\ref{lowest order}) agrees with
(\ref{two-point function}) in the limit (\ref{limit}).
For simplicity, we first take the $n\rightarrow\infty$ limit.
Then, (\ref{lowest order}) can be evaluated as
\begin{align}
\frac{g_{\rm PW}^4k^2}{\mu^2n^2\nu}\int_{\mu|t-t'|}^{\infty}ds\sum_{I,J=1}^{\nu}
\sum_{j=\frac{1}{2}|I-J|}^{\infty}e^{-(2j+1)s}
&=\frac{g_{\rm PW}^4k^2}{\mu^2n^2\nu}\int_{\mu|t-t'|}^{\infty}ds \frac{e^{-s}}{1-e^{-2s}}\sum_{I,J=1}^{\nu}
e^{-|I-J|s} \nonumber\\
&=\frac{g_{\rm PW}^4k^2}{\mu^2n^2}\int_{\mu|t-t'|}^{\infty}ds \frac{e^{-s}}{(1-e^{-s})^2} \nonumber\\
&=\frac{g_{\rm PW}^4k^2}{\mu^2n^2}\frac{e^{-\mu|t-t'|}}{1-e^{-\mu|t-t'|}}
\ .
\label{calculation of two-point function}
\end{align}
By using (\ref{'t Hooft coupling}), we find
that the last expression indeed agrees with (\ref{two-point function}).

Similarly, one can calculate the three-point and
four-point functions corresponding
to those in (\ref{four-point function}) in the
free theory.
The results are
\begin{align}
&\frac{k}{n^3\nu}
\Bigl\langle{\cal O}_{45}(t_1){\cal O}_{56}(t_2){\cal O}_{64}(t_3)
\Bigr\rangle_{\rm PW,free}
 \! = \! \frac{g_{\rm PW}^6k^3}{4\mu^3n^3\nu}
\!
\sum_{I,J=1}^{\nu}
\!
\sum_{j=\frac{1}{2}|n_I-n_J|}^{\frac{1}{2}(n_I+n_J)-1}
\!\!\!
\frac{e^{-\mu(j+\frac{1}{2})(|t_1-t_2|+|t_2-t_3|+|t_3-t_1|)}}{(j+\frac{1}{2})^2} \ , \nonumber\\
&\frac{k^2}{n^4\nu}\Bigl\langle {\cal O}_{45}(t_1){\cal O}_{56}(t_2){\cal O}_{67}(t_3)
{\cal O}_{74}(t_4)\Bigr\rangle_{\rm PW,free}  \nonumber\\
&=\frac{g_{\rm PW}^8k^4}{8\mu^4n^4\nu}
\sum_{I,J=1}^{\nu}\sum_{j=\frac{1}{2}|n_I-n_J|}^{\frac{1}{2}(n_I+n_J)-1}
\frac{e^{-\mu(j+\frac{1}{2})(|t_1-t_2|+|t_2-t_3|+|t_3-t_4|+|t_4-t_1|)}}{(j+\frac{1}{2})^3} \ .
\label{four-point function in PWMM}
\end{align}
Similarly to (\ref{calculation of two-point function}),
one can show that
the three-point and four-point functions
in (\ref{four-point function in PWMM}) agree
with those in (\ref{four-point function}) in the limit (\ref{limit}).
%
%
Thus we have confirmed that (\ref{correspondence})
holds in the free theory case.



\begin{figure}[t]
\begin{center}
   \includegraphics[width=8cm]{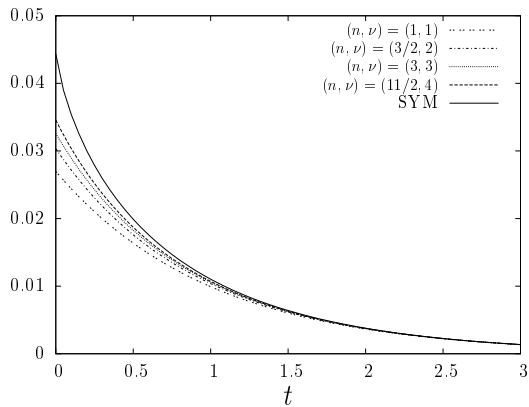}
\end{center}
\caption{The four-point function (\ref{four-point function in PWMM})
for the PWMM in the free theory case with $\mu=2$ is plotted
for backgrounds (\ref{our background})
with $(n,\nu)=(1,1)$, $(\frac{3}{2},2)$, $(3,3)$, $(\frac{11}{2},4)$.
The normalization constant in (\ref{four-point function})
is chosen such that it formally corresponds to $\lambda=16$.
We observe clear convergence towards the result
(\ref{four-point function}) for ${\cal N}=4$ SYM
in the free theory case represented by the solid line.
}
\label{fourpt}
\end{figure}

Let us discuss the rate of convergence.
As an example, we consider
the four-point function
in the PWMM given by (\ref{four-point function in PWMM}),
and see how it
converges to that in
${\cal N}=4$ SYM given by (\ref{four-point function})
in the free theory case.
Here we take the limit (\ref{limit}) with
a choice $n=\frac{1}{2}\nu^2-\nu+ \frac{3}{2}$, which amounts to
considering a sequence
\begin{align}
(n,\nu)=(1,1), \  \left(\frac{3}{2},2\right),
\  (3,3), \ \left(\frac{11}{2},4\right), \cdots \ .
\end{align}
In figure \ref{fourpt} we plot
the results for this sequence, which show clear convergence
towards the result for ${\cal N}=4$ SYM.
In our Monte Carlo simulation, which will be discussed
in the next section, we take the background $(n,\nu)=(\frac{3}{2},2)$ in this
sequence.

\section{Monte Carlo method}
\label{sec:monte}

In this section we discuss our Monte Carlo method for
studying ${\cal N}=4$ SYM.
Due to
the large-$N$ reduction,
${\cal N}=4$ SYM on $R\times S^3$
is equivalent
to the PWMM (\ref{pp-action}),
which we actually simulate.
Since
the PWMM is a one-dimensional theory,
we need to introduce an IR and UV cutoffs in the $t$-direction.
The IR cutoff is introduced
by compactifying the $t$-direction to a circle\footnote{Strictly speaking,
SUSY is softly broken for finite $\beta$
since the SUSY transformation for the PWMM is $t$-dependent
unlike in the D0-brane system corresponding to the
$\mu=0$ case.}
of circumference $\beta$
with
periodic boundary conditions
on both scalars $X_{i}(t)$ and fermions $\Psi_{\alpha}(t)$.
%
Following refs.~\cite{AHNT,Hanada:2008gy,Hanada:2008ez,%
Hanada:2009ne,Hanada:2011fq},
we introduce a sharp UV cutoff in the Fourier space
after fixing the gauge completely.
This is possible since the $t$-direction is one-dimensional.

First we take the static diagonal gauge
\begin{equation}
A(t) = \frac{1}{\beta} \, {\rm diag} (\alpha_{1},\cdots ,\alpha_{N}) \ ,
\end{equation}
where $\alpha_{a}$ are constant in time.
This condition does not fix the gauge completely,
and in fact there is a residual gauge symmetry
\begin{equation}
\alpha_{a}\mapsto \alpha_{a}+2\pi\nu_{a} \ ,\quad
\tilde{X}_{i,n}^{ab} \mapsto \tilde{X}_{i,n-\nu_{a}+\nu_{b}}^{ab} \ ,\quad
\tilde{\Psi}_{\alpha ,n}^{ab} \mapsto
\tilde{\Psi}_{\alpha ,n-\nu_{a}+\nu_{b}}^{ab} \ ,
\end{equation}
where $\nu_{a}$ is an integer.
This residual symmetry can be fixed by imposing
$-\pi <\alpha_{a}\leq \pi$.
Corresponding to the above gauge choice,
the Faddeev-Popov term
\begin{equation}
S_{\rm FP} = -\sum_{a<b}2\ln{\left|
\sin{\frac{\alpha_{a}-\alpha_{b}}{2}} \right|}
\end{equation}
should be added in the action.
Then we introduce a cutoff $\Lambda$
in the Fourier-mode expansion
\begin{equation}
X_{i}(t) = \sum_{n=-\Lambda}^{\Lambda} \tilde{X}_{i,n}e^{i\omega nt} \ ,
\quad \omega\equiv \frac{2\pi}{\beta} \ ,
\end{equation}
and similarly for the fermions \cite{Hanada:2007ti}.
Since there are no UV divergences in the one-dimensional theory,
one can obtain the original PWMM by taking the limits
$\beta\rightarrow\infty \ ,
\frac{\Lambda}{\beta}\rightarrow \infty $.

The integration over fermionic variables
yields a Pfaffian, which is complex in general.
As is done in the previous works \cite{AHNT,Hanada:2008gy,%
Hanada:2008ez,Hanada:2009ne,Hanada:2011fq}
on the D0-brane system,
we simply take the absolute value of the Pfaffian
assuming that the phase does not affect the
results.\footnote{See appendix C of ref.~\cite{Hanada:2011fq}
for possible justification.
Ref.~\cite{Catterall:2010gf}
shows that the phase of the Pfaffian is negligible
in Monte Carlo simulation of the lattice regularized PWMM
with the trivial background ($n=\nu =1,k=N=3$)
at finite temperature.}
The model obtained in this way
can be simulated by the Rational Hybrid Monte Carlo (RHMC)
algorithm \cite{Clark:2003na}.
This method has been applied extensively
to the D0-brane system corresponding to $\mu=0$,
and the results confirmed the gauge/gravity duality
for various observables \cite{AHNT,Hanada:2008gy,%
Hanada:2008ez,Hanada:2009ne,Hanada:2011fq}.\footnote{See
refs.\ \cite{Catterall:2007fp,Catterall:2008yz,Catterall:2009xn,Kadoh:2012bg}
for Monte Carlo calculations
based on the lattice regularization.
Our method has also been applied to other
SUSY matrix quantum mechanics with less supercharges and
their nonperturbative properties have been studied \cite{Hanada:2010jr}.
}
See appendix B
of ref.~\cite{Hanada:2011fq} for the details of the algorithm.

We start our simulation with an initial configuration
given by (\ref{our background})
with the parameters $(n,\nu ,k)=(\frac{3}{2},2,2)$,
which corresponds to the matrix size $N=6$.
Since the parameter $\mu$ in the action (\ref{pp-action})
can be scaled away by appropriate redefinition of the fields
and the parameters,
we take $\mu=2$ without loss of generality,\footnote{This
convention is different from the one in
ref.~\cite{Honda:2010nx}, where we set the 't Hooft coupling
$g_{\rm PW}^{2}N$ to unity analogously to the studies of the
D0-brane system \cite{AHNT,%
Hanada:2008gy,Hanada:2008ez,Hanada:2009ne,Hanada:2011fq}.
The dictionary between the two conventions is given by
\beqa
&~& X_{M}^\prime = (g_{\rm PW}^2 N)^{-1/3}X_{M} \ , \quad
A_{0}^\prime = (g_{\rm PW}^2 N)^{-1/3}A_{0} \ ,\quad
\psi^\prime = (g_{\rm PW}^2 N)^{-1/2}\psi \ ,
\nonumber \\
&~& t^\prime = (g_{\rm PW}^2 N)^{1/3}t \ , \quad
g_{\rm PW}^2 N =\left( \frac{\mu^\prime}{2}\right)^{-3} \ ,
\eeqa
where
variables with (without) a prime
correspond to the previous (present) convention, respectively.
In particular, the values of $\mu'$ in the previous convention are
$4.0$, $2.0$, $1.3$ for $\lambda=0.55$, $4.39$, $16.0$, respectively.
\label{footnote:convention}
}
which corresponds to a unit sphere $S^{3}$ due to
(\ref{R-mu-rel}).
%
Then, for the chosen background
$(n,\nu ,k)=(\frac{3}{2},2,2)$,
the relation
between $g_{\rm PW}^2N$ and $\lambda$
becomes
\begin{align}
\lambda =\frac{4}{9} \pi^2 (g_{\rm PW}^2 N) \ ,
\label{g-lambda-rel}
\end{align}
due to (\ref{limit}) and (\ref{'t Hooft coupling}).
We choose the
coupling constant of the PWMM as
$g_{\rm PW}^{2}N=0.125$, $1.0$, $3.64$,
which correspond to
the coupling constant of $\mathcal{N}=4$ SYM
given by $\lambda=0.55$, $4.39$, $16.0$,
respectively,
according to (\ref{g-lambda-rel}).
For these values of $\lambda$,
we find that transitions to other vacua
do not occur during the simulation
as we discuss in appendix \ref{appendix:C}.
The IR cutoff in the $t$-direction is taken as
$\beta=10.0$, $5.0$, $3.25$, respectively, for each
couplings,\footnote{This choice of $\beta$ corresponds to
taking $\beta=5.0$ in our previous convention
in footnote \ref{footnote:convention},
which is considered to be large enough for calculating
correlation functions
as analogous studies in
the D0-brane system indicate \cite{Hanada:2009ne,Hanada:2011fq}.
\label{footnote:beta-choice}
}
while the UV cutoff parameter $\Lambda$ in the $t$-direction
is taken as $\Lambda =12$ for all cases.

The dependence of our results on the regularization parameters
is discussed in appendix \ref{appendix:D}.
In particular, we have checked that our results for the two-point
functions after taking the ratio to the free theory case
do not change significantly for larger $n$, $\nu$ and $k$.
Also we find that finite-$\Lambda$ effects are negligible,
and hence the SUSY breaking by such effects can be safely ignored.

\section{Results}
\label{sec:results}

In this section we present our results
for the correlation functions of CPOs in $\mathcal{N}=4$ SYM.
These results are obtained by
using the relationship (\ref{correspondence})
and calculating the corresponding correlation functions
in the PWMM by the Monte Carlo method described
in the previous section.

\begin{figure}[t]
\begin{center}
   \includegraphics[width=7.2cm]{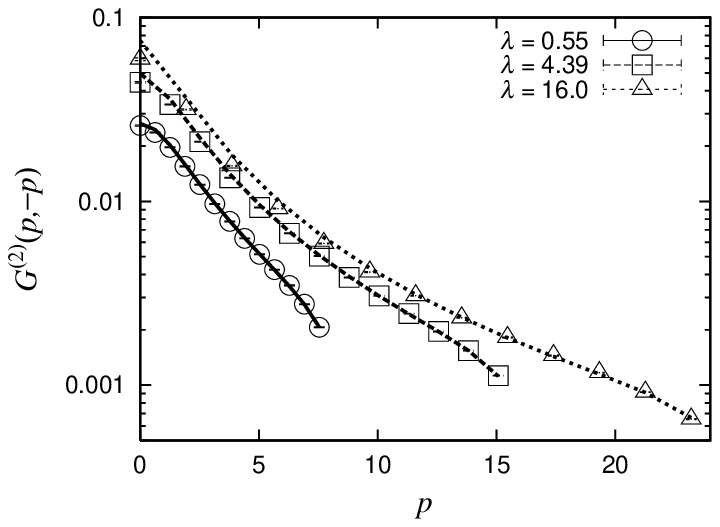}
   \includegraphics[width=7.2cm]{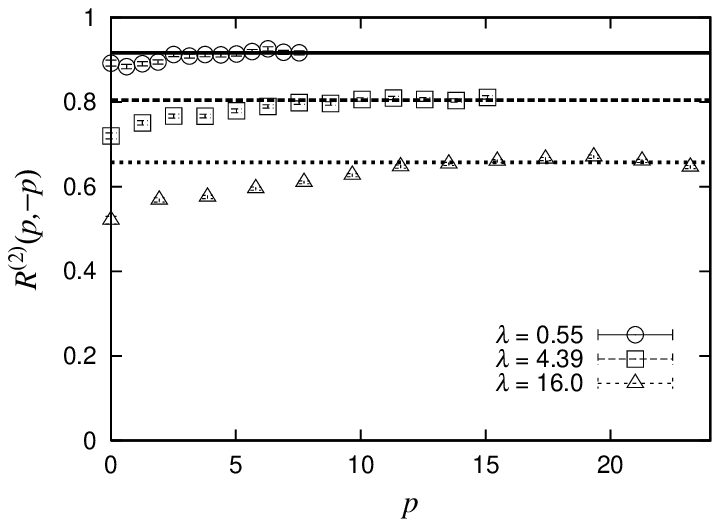}
\end{center}
\caption{
(Left) The two-point function $G^{(2)}(p,-p)$
is plotted as a function of $p$ in the log scale
for $\lambda= 0.55$, $4.39$, $16.0$.
The curves represent the corresponding free theory results
multiplied by the values obtained as the height of the plateau
in the right panel.
%
(Right) The ratio $R^{\rm (2)}(p,-p)$ of the two-point function
to the corresponding free theory result is plotted
for
$\lambda= 0.55$, $4.39$, $16.0$.
The horizontal lines represent the fits to the plateau behavior,
which give estimates for $c^{(2)}$.
}
\label{figures-2pt}
\end{figure}

In figure\ \ref{figures-2pt} (Left) we present our results for
the two-point function\footnote{The normalization of the correlation
functions (\ref{G2-def}), (\ref{G3-def}), (\ref{G4-def})
we adopt in this paper is a natural one from the viewpoint
of the PWMM. The 't Hooft coupling $g_{\rm PW}^2 N$ in the denominator
comes from the rescaling of the scalar fields which is needed to
make the kinetic term in the action (\ref{pp-action}) canonical,
while the $\mu$ in the numerator is introduced to
make the correlation functions dimensionless.
The adopted normalization is different from the one on the right-hand side
of eq.~(\ref{correspondence}), which becomes finite
in the limit (\ref{limit}).
This issue is irrelevant, however, when we take the ratio
to the free theory case in (\ref{ratio-2pt}),
(\ref{ratio-3pt}) and (\ref{ratio-4pt})
since the common factors cancel.
}
\begin{align}
G^{(2)}(p,-p)
&=
\left( \frac{\mu}{g_{\rm PW}^{2}N} \right)^{2} \left\langle
 \widetilde{O}_{45}(p) \,  \widetilde{O}_{54}(-p)
\right\rangle_{\rm PW}
\ ,
\label{G2-def}
\end{align}
where
we have defined
the Fourier transform of an operator ${\cal O}(t)$ as
\beq
\widetilde{\cal O}(p)=\frac{1}{\beta} \int_{0}^{\beta}dt \,
{\cal O}(t) \, e^{-ipt} \ .
\eeq
In order to study the non-renormalization property,
we compare our data
with the free theory results, which are calculated analytically
by just switching off the interaction terms
in the reduced model with the same regularization parameters.
(See appendix \ref{appendix:F} for explicit results
not only for two-point functions but also for
three-point and four-point functions.)
In figure\ \ref{figures-2pt} (Right) we plot the ratio
\begin{equation}
R^{\rm (2)}(p,-p)=
\frac{\Big\langle
 \widetilde{O}_{45}(p) \,  \widetilde{O}_{54}(-p)
 \Big\rangle_{\rm PW} }
     {\Big\langle
 \widetilde{O}_{45}(p) \,  \widetilde{O}_{54}(-p)
 \Big\rangle_{\rm PW,free}} \ .
\label{ratio-2pt}
\end{equation}
The momentum dependence of the correlation function is almost canceled
by taking the ratio, and we observe a nice plateau behavior.
By fitting
$R^{\rm (2)}(p,-p)$ to a constant in the momentum region
$p=\frac{2\pi n }{\beta}$, where $6 \le n \le \Lambda \equiv 12$,
we obtain $0.917(2)$, $0.805(2)$, $0.658(3)$
for $\lambda=0.55$, $4.39$, $16.0$, respectively.
In figure\ \ref{figures-2pt} (Left) we also plot the free theory results
multiplied by the overall constants obtained in this way.
Thus we find that our data are in good agreement with the
corresponding free theory result
up to an overall constant depending on $\lambda$.
This is remarkable considering
that the value of
the two-point function changes by orders of magnitude
as a function of $p$.

As we have reviewed in section \ref{sec:corr-fun-sym},
the form of the two-point function is fixed
by the conformal invariance of $\mathcal{N}=4$ SYM,
and therefore
the ratio $R^{\rm (2)}(p,-p)$ should become a constant,
which corresponds to $c^{(2)}$ in
(\ref{CPO-cor-def-2pt})
in the large-$N$ limit (\ref{limit}).
The height of the plateau
in figure\ \ref{figures-2pt} (Right)
gives an estimate for $c^{(2)}$ with the present matrix size,
which decreases from 1 as $\lambda$ increases.
On the other hand,
the SUSY non-renormalization property
(\ref{non-renormalization of 3-pt fn})
implies
%
that $c^{(2)}=1$.
We therefore consider that the
height of the plateau approaches 1 for any $\lambda$
as we take the limit (\ref{limit}).

%

\begin{figure}[t]
\begin{center}
   \includegraphics[width=7.2cm]{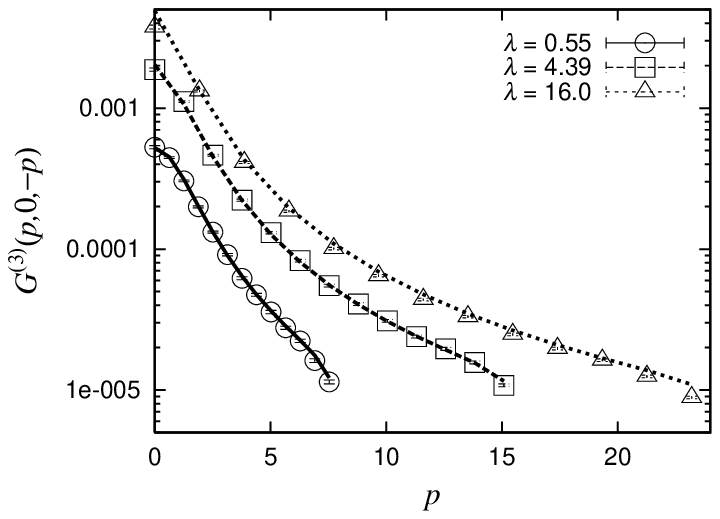}
   \includegraphics[width=7.2cm]{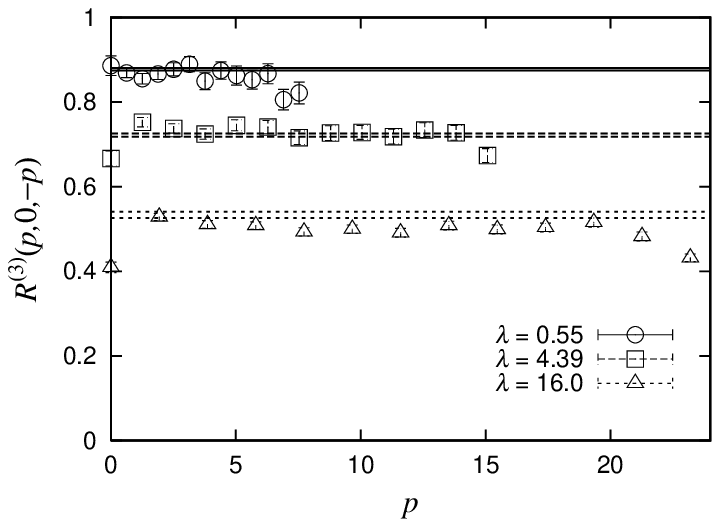}
\end{center}
\caption{
(Left) The three-point function $G^{(3)}(p,0,-p)$
is plotted as a function of $p$ in the log scale
for $\lambda= 0.55$, $4.39$, $16.0$.
The curves represent the corresponding free theory results
multiplied by $(c^{(2)})^{3/2}$ using $c^{(2)}$ obtained
from figure~\ref{figures-2pt} for each $\lambda$.
(Right) The ratio $R^{\rm (3)}(p,0,-p)$
of the three-point function
to the corresponding free theory result is plotted
for
$\lambda= 0.55$, $4.39$, $16.0$.
The horizontal line represents
$(c^{(2)})^{3/2}$ using $c^{(2)}$ obtained from
figure~\ref{figures-2pt}.
}
\label{figures-3pt}
\end{figure}

Let us move on to the three-point function
\begin{equation}
G^{(3)}(p,0,-p) =
\left( \frac{\mu}{g_{\rm PW}^{2}N} \right)^{3}
\Big\langle \tilde{{\cal O}}_{45}(p)\tilde{{\cal O}}_{56}(0)
\tilde{{\cal O}}_{64}(-p)\Big\rangle_{\rm PW} \ ,
\label{G3-def}
\end{equation}
which is shown in figure \ref{figures-3pt} together with
the ratio
\begin{equation}
R^{\rm (3)}(p,0,-p)=
\frac{\Big\langle \tilde{{\cal O}}_{45}(p)
 \tilde{{\cal O}}_{56}(0)
\tilde{{\cal O}}_{64}(-p)\Big\rangle_{\rm PW}}
 {\Big\langle \tilde{{\cal O}}_{45}(p)
\tilde{{\cal O}}_{56}(0)
\tilde{{\cal O}}_{64}(-p)\Big\rangle_{\rm PW,free}} \ .
\label{ratio-3pt}
\end{equation}
We observe that the three-point function agrees
with the free theory results up to an overall constant.
Since the form of the three-point function is determined
by the conformal symmetry as in the case of two-point functions,
the ratio $R^{\rm (3)}(p,0,-p)$ should become a constant,
which corresponds to $c^{(3)}$ in (\ref{CPO-cor-def-3pt}),
in the limit (\ref{limit}).
The height of the plateau in figure \ref{figures-3pt} (Right)
gives an estimate for $c^{(3)}$ with the present matrix size.

The AdS/CFT correspondence
predicts (\ref{def-c2}) and (\ref{def-c3}), which implies
\beq
c^{(3)}=(c^{(2)})^{3/2} \ .
\label{2pt-3pt-rel}
\eeq
In order to see
whether our data is consistent with (\ref{2pt-3pt-rel}),
we use the value of $c^{(2)}$ extracted from
our results for two-point functions, and plot $(c^{(2)})^{3/2}$
in figure~\ref{figures-3pt} (Right).
The margin between the lines represents the fitting error
in estimating $c^{(2)}$ from the plateau height.
Our results for the ratio $R^{\rm (3)}(p,0,-p)$
are in reasonable agreement with
$(c^{(2)})^{3/2}$ obtained from the two-point function,
which implies
consistency with
(\ref{2pt-3pt-rel}).
Thus we find that
the weaker form of the SUSY non-renormalization property
holds even at the regularized level.
In the large-$N$ limit (\ref{limit}),
both sides of (\ref{2pt-3pt-rel}) are expected to approach 1.


%
\begin{figure}[t]
\begin{center}
\includegraphics[width=7.2cm]{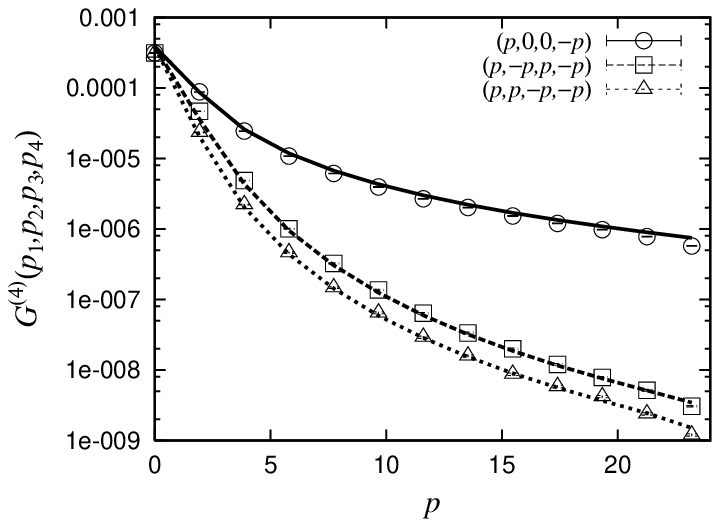}
\includegraphics[width=7.2cm]{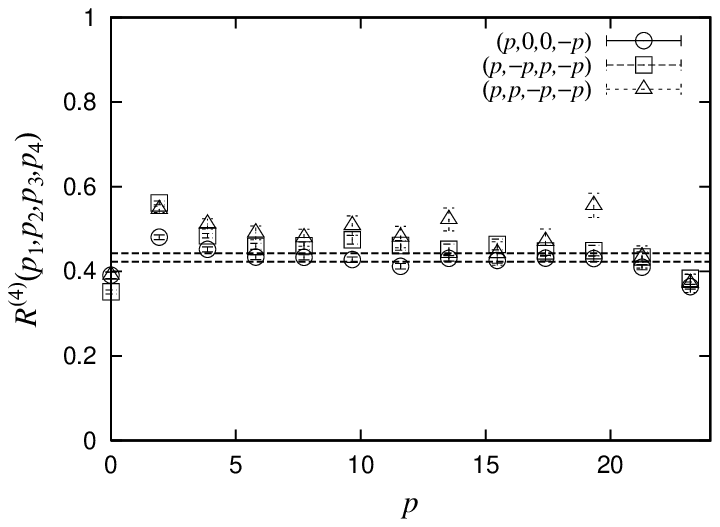}
\end{center}
\caption{(Left) The four-point function $G^{(4)}(p_1,p_2,p_3,p_4)$
is plotted for three types of momentum configuration
$(p_1,p_2,p_3,p_4)=(p,0,0,-p),~(p,-p,p,-p),~(p,p,-p,-p)$
as a function of $p$
in the log scale for $\lambda=16.0$.
The curves represent the corresponding free theory results
multiplied by $(c^{(2)})^{2}$ using $c^{(2)}$ obtained from
figure~\ref{figures-2pt}.
(Right) The ratio $R^{\rm (4)}(p_1,p_2,p_3,p_4)$
of the four-point function
to the corresponding free theory result is plotted
for three types of momentum configuration
$(p_1,p_2,p_3,p_4)=(p,0,0,-p),~(p,-p,p,-p),~(p,p,-p,-p)$
as a function of $p$
for $\lambda= 16.0$.
The horizontal line represents
$(c^{(2)})^{2}$ using $c^{(2)}$ obtained from
figure~\ref{figures-2pt}.
}
\label{figures-4pt}
\end{figure}

Finally we discuss our results for the four-point functions
\begin{equation}
G^{(4)}(p_1,p_2,p_3,p_4) =
\left( \frac{\mu}{g_{\rm PW}^{2}N} \right)^{4}
\Big\langle \tilde{{\cal O}}_{45}(p_1)\tilde{{\cal O}}_{56}(p_2)
\tilde{{\cal O}}_{67}(p_3) \tilde{{\cal O}}_{74}(p_4)
\Big\rangle_{\rm PW} \ .
\label{G4-def}
\end{equation}
Since the four-point functions we study
are neither extremal nor next-to-extremal,
the SUSY non-renormalization property can be violated.
Indeed the AdS/CFT correspondence predicts
the explicit form of the violation
given by (\ref{ads-cft-pred-4pt}).
The prediction of the AdS/CFT correspondence is obtained
in the strong coupling limit,
and the violation
is expected to become smaller as the coupling constant decreases.
Therefore, here we
focus on the case of
the largest coupling constant $\lambda= 16.0$.
In figure~\ref{figures-4pt} (Left)
we plot our results for the four-point functions with three types of
momentum configuration
$(p_1,p_2,p_3,p_4)=(p,0,0,-p)$, $(p,-p,p,-p)$ and $(p,p,-p,-p)$.
In figure~\ref{figures-4pt} (Right)
we plot the ratio $R^{\rm (4)}$ of the four-point function
to the free theory result
\begin{align}
&R^{\rm (4)}(p_1,p_2,p_3,p_4)
=\frac{\Big\langle \tilde{{\cal O}}_{45}(p_1)
\tilde{{\cal O}}_{56}(p_2)
\tilde{{\cal O}}_{67}(p_3)
\tilde{{\cal O}}_{74}(p_4)\Big\rangle_{\rm PW}}
{\Big\langle \tilde{{\cal O}}_{45}(p_1)
\tilde{{\cal O}}_{56}(p_2)
\tilde{{\cal O}}_{67}(p_3)
\tilde{{\cal O}}_{74}(p_4) \Big\rangle_{\rm PW,free}}
\label{ratio-4pt}
\end{align}
for each momentum configuration
as a function of $p$.

\begin{figure}[t]
\begin{center}
\includegraphics[width=7.2cm]{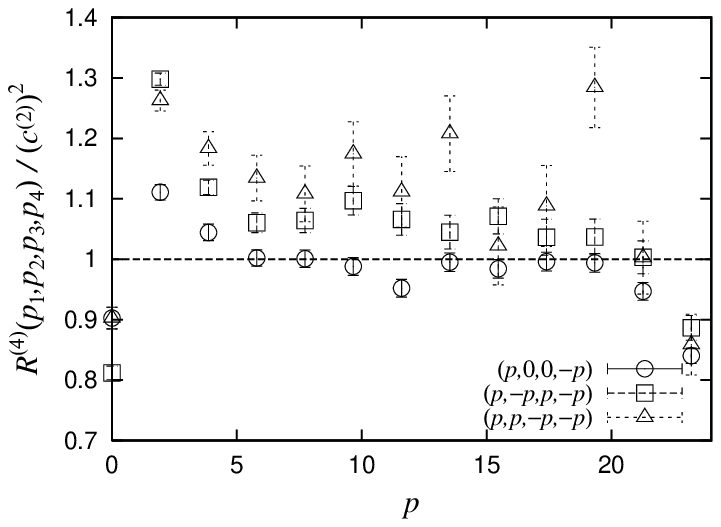}
\includegraphics[width=7.2cm]{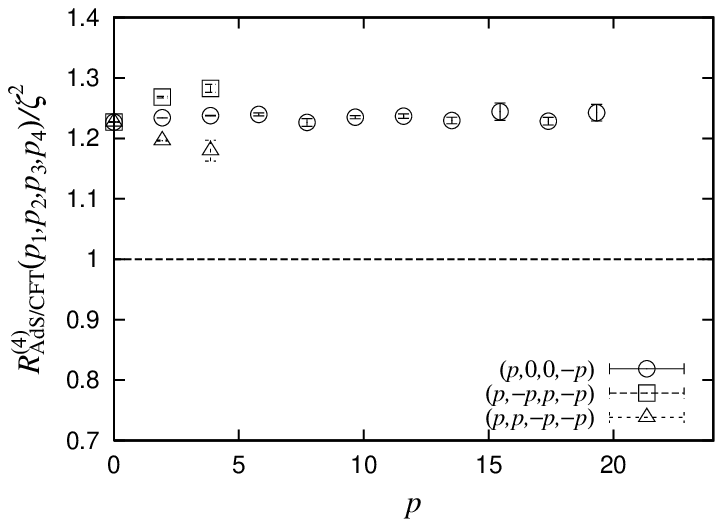}
\end{center}
\caption{(Left) The normalized ratio
$R^{\rm (4)}(p_1,p_2,p_3,p_4)/(c^{\rm (2)})^2$ is plotted
for three types of momentum configuration
$(p_1,p_2,p_3,p_4)=(p,0,0,-p),~(p,-p,p,-p),~(p,p,-p,-p)$
as a function of $p$
for $\lambda = 16.0$.
%
(Right) The normalized ratio of the four-point functions
$R^{\rm (4)}_{\rm AdS/CFT}(p_1,p_2,p_3,p_4)/\zeta^2$
obtained from the AdS/CFT correspondence
is plotted for three types of momentum configuration
$(p_1,p_2,p_3,p_4)=(p,0,0,-p),~(p,-p,p,-p),~(p,p,-p,-p)$
as a function of $p$.
}
\label{figures-4pt_2}
\end{figure}

The weaker form of the non-renormalization property implies
$R^{\rm (4)}(p_1,p_2,p_3,p_4)=(c^{(2)})^2$.
In order to see its violation,
we plot $(c^{(2)})^2$
in figure \ref{figures-4pt} (Right)
using the value of $c^{(2)}$
obtained from our results for the two-point function.
The margin between the lines
represents the fitting error in estimating $c^{(2)}$
from
the plateau height in figure\ \ref{figures-2pt} (Right).
We find that our data for $R^{\rm (4)}(p_1,p_2,p_3,p_4)$
appear systematically larger
than the value of $(c^{(2)})^2$
in sharp contrast to our results for the three-point function
shown in figure~\ref{figures-3pt} (Right).
In figure~\ref{figures-4pt_2} (Left)
we plot the ratio $R^{\rm (4)}/ (c^{(2)})^2$ obtained by
Monte Carlo data.
We find that
most of the data points lie within the range $1 \sim 1.3$,
which suggests the violation of the non-renormalization property.

Let us then see whether this violation suggested from our
Monte Carlo data is consistent with
the prediction of the AdS/CFT correspondence.
For that we need to translate (\ref{ads-cft-pred-4pt})
into the four-point functions
we measure directly in the simulation as
\begin{eqnarray}
&& \Big\langle \hat{{\cal O}}_{45}(p_1)
\hat{{\cal O}}_{56}(p_2)
\hat{{\cal O}}_{67}(p_3)
\hat{{\cal O}}_{74}(p_4)\Big\rangle \times
(2 \pi)^4 \delta^{(4)} (p_1 + \ldots + p_4)
\nonumber \\
&=& \frac{\lambda^4}{2^8 \pi^8 N^2}
\int \left(\prod_{i=1}^4
dt_{i} \frac{d\Omega_3^{(i)}}{2\pi^2}  \right)
e^{-i \sum_i^4 p_i t_i}
e^{2 \sum_i^4 t_i}
\frac{\zeta ^2 c(u,v)}{x_{12}^2 x_{23}^2 x_{34}^2 x_{41}^2} \ ,
\label{4pt-measured}
\end{eqnarray}
where we have defined
\begin{align}
\hat{{\cal O}}_{ab}(p)=
\int dt \, \bar{{\cal O}}_{ab}(t) \, e^{-ipt} \ .
\label{hatO-def}
\end{align}
The factor $\exp(2 \sum_i^4 t_i)$ in (\ref{4pt-measured})
comes from the transformation (\ref{scalar_trans})
associated with the conformal map from $R^{4}$ to $R\times S^{3}$.
The corresponding free theory result
can be obtained by setting $c(u,v)=1$ and $\zeta=1$
in (\ref{4pt-measured})
as one can see from (\ref{4-point function}) and (\ref{free-coeff-c}).
Then the ratio of the four-point function
to the corresponding free theory result for
$\mathcal{N}=4$ SYM is predicted from the AdS/CFT correspondence
as
\begin{align}
R^{\rm (4)}_{\rm AdS/CFT}(p_1,p_2,p_3,p_4)
&=\frac{\Big\langle \hat{{\cal O}}_{45}(p_1)\hat{{\cal O}}_{56}(p_2)
\hat{{\cal O}}_{67}(p_3)
\hat{{\cal O}}_{74}(p_4) \Big\rangle}
      {\Big\langle
\hat{{\cal O}}_{45}(p_1)
\hat{{\cal O}}_{56}(p_2)
\hat{{\cal O}}_{67}(p_3)
\hat{{\cal O}}_{74}(p_4)\Big\rangle_{\rm free}}
\nonumber
\\
& = \zeta ^2 F(p_1 , p_2 , p_3 , p_4 ) \ ,
\label{RATIO4pt}
\end{align}
where $F(p_1 , p_2 , p_3 , p_4 )$ can be calculated by
(\ref{4pt-measured}) and the corresponding expression for the
free theory.
Figure~\ref{figures-4pt_2} (Right)
shows
the function $F(p_1 , p_2 , p_3 , p_4 ) = R^{\rm (4)}_{\rm AdS/CFT}/\zeta^2$
obtained in the way described in appendix \ref{appendix:B}.
%
Thus the AdS/CFT correspondence
predicts $R^{\rm (4)}_{\rm AdS/CFT}/\zeta^2 = 1.2 \sim 1.3$,
which roughly agrees with the violation of the non-renormalization
property observed on the gauge theory side.
In fact the agreement
for the momentum configuration $(p,p,-p,-p)$ is remarkable;
the gauge theory side and the gravity side both predict
a value around 1.2.

For other momentum configurations,
the gauge theory results turn out to be approximately
20\% smaller than
expected from the AdS/CFT correspondence.
We consider that this is
due to
the finite IR
cutoff effects in the
$t$-direction.\footnote{As another possible source of discrepancies,
we note that our data for the four-point function are
obtained for $\lambda=16.0$,
while the prediction of the AdS/CFT correspondence is obtained for
$\lambda=\infty$.}
For instance, figure \ref{figures-2pt} (Right)
shows that our data in the small $p$ region are smaller
than the expected plateau behavior
by roughly 10--20\%
for $\lambda=16.0$.
Also figure \ref{figures-3pt} (Right) shows that
the ratio for the three-point function turns out to be
several \% smaller than expected from
the weaker form of the non-renormalization property.
This might be attributed to the fact that the three-point function
we consider involves $p=0$ as one of its arguments.
The four-point function for
the momentum configuration $(p,0,0,-p)$ may well be subject to such
artifacts.
It is also conceivable that the four-point function for
the momentum configuration $(p,-p,p,-p)$
is subject to such IR artifacts
since one obtains
zero momentum
whenever two adjacent momenta are added in planar diagrams, which
dominate at large $N$.

\section{Summary and discussions}
\label{sec:summary}

In this paper we have studied
${\cal N}=4$ SYM by Monte Carlo simulation,
and calculated, in particular,
the correlation functions of chiral primary operators
to test the predictions of
the AdS/CFT correspondence in the strong coupling limit.
Our results for the three-point function
turn out to be consistent with the
non-renormalization property in the weaker form,
while our results for the four-point function
suggest its violation.
These results
are consistent with
the predictions from the AdS/CFT correspondence.
In particular, the violation of the
non-renormalization property observed in the four-point function
has the same orders of magnitude
as predicted
by the AdS/CFT correspondence.

This is remarkable considering the rather small system size,
which is represented by $N=6$ and $\Lambda=12$ in our simulation.
The number of degrees of freedom is $(N^2-1)(2\Lambda+1)=875$,
which roughly corresponds to the SU(2) gauge theory
on the $4^4$ lattice.
The crucial point was to use the idea of
the large-$N$ reduction, which enables us to regularize
${\cal N}=4$ SYM respecting 16 SUSYs.
In particular,
the finite UV cutoff effects
in the raw data for the correlation functions
turned out to be almost canceled up to an overall constant
by taking the ratio to the results for the free theory
with the same regularization parameters.
Furthermore, our results suggest that
the possible finite UV cutoff effects in the overall constant
for the correlation functions
can be mostly absorbed by
appropriate normalization of the operators.
These features are considered a big advantage of our approach,
which made it possible to test the predictions of the
AdS/CFT correspondence with
the available computer resources.

In fact field theoretical analyses in ${\cal N}=4$ SYM suggest
that the two-point and three-point functions
are not renormalized including the overall constant factor.
This is a stronger statement than the
prediction of the AdS/CFT correspondence
for the two-point and three-point functions.
The overall constants
extracted from our Monte Carlo data,
however, show some deviation from this statement
as one increases the coupling constant.
We consider it likely that this deviation is due to the finite
UV cutoff effects and that it will disappear if one takes
the limit (\ref{limit}) to infinite matrix size.
Since our data are already
consistent with the non-renormalization
property in the weaker form, we only need to confirm that
the overall constant for the two-point function approaches
unity as one increases the matrix size
for arbitrary coupling constant.
We leave this issue for future investigations.


Now that we have established a new method for
nonperturbative calculations in ${\cal N}=4$ SYM
in the large-$N$ limit,
it would be interesting to apply the method to various other quantities.
We are currently working on the calculation of Wilson loops.
Preliminary results for the circular
Wilson loops are already
reported in ref.~\cite{Nishimura:2009xm,Honda:2011qk},
where one can see reasonable agreement with
the analytic results \cite{Erickson:2000af,Pestun:2007rz}.
We are going to extend the calculation to non-BPS Wilson loops,
which cannot be calculated analytically,
and to test the prediction \cite{maldloop,PML}
from the AdS/CFT correspondence.

It would also be interesting to study
SYM in other dimensions in a similar way.
For instance, it would be interesting to study
three-dimensional $\mathcal{N}=8$ SYM
on
$R\times S^2$
with
16 SUSYs and to test the gauge/gravity
correspondence.
The theory has many classical vacua, all of which
preserve the full SUSYs and have
a known dual gravity description \cite{Lin:2005nh}.
In order to test the more standard
AdS/CFT correspondence associated
with the D2-brane \cite{Kanitscheider:2008kd},
one has to study the SYM on
$R^3$
instead.
For that, we first consider
the SYM on
$R\times S^2$,
and send the radius of
$S^2$
to infinity.
This is a slight complication compared with
${\cal N}=4$ SYM studied in this paper,
where
one can use the conformal invariance
to map the theory on
$R\times S^3$
to the one
on
$R^4$
and vice versa.
%

To conclude, we consider it very interesting that
the planar large-$N$ limit allows us to regularize gauge theories
preserving 16 SUSYs.
This not only enables us to study the SUSY theories
by Monte Carlo simulation without fine-tuning,
but also enables us to access interesting physics already
with rather small system size as the present work clearly demonstrates.



\acknowledgments

We would like to thank H.~Kawai and Y.~Kitazawa for valuable discussions.
The computations were carried out
on the B-factory computer system at KEK.
The work of M.~H.\ is supported
by  Japan Society for the Promotion of Science (JSPS).
The work of G.~I.\ and S.-W.~K.\
is supported by the National Research Foundation
of Korea (NRF) Grant funded by the Korean Government
(MEST 2005-0049409 and NRF-2009-352-C00015).
The work of J.~N.\ and A.~T.\ is supported
by Grant-in-Aid for Scientific
Research
(No.\ 20540286, 24540264, and 23244057)
from JSPS.

\appendix

\section{Four-point function predicted by the AdS/CFT correspondence}
\label{appendix:A}

The prediction for the four-point function
in ${\cal N}=4$ SYM on
$R^4$
is obtained
from the AdS/CFT correspondence
as (\ref{ads-cft-pred-4pt}) in
refs.~\cite{Arutyunov:2000py,Arutyunov:2000ku,Arutyunov:2002fh}.
The function $c(u,v)$ in (\ref{ads-cft-pred-4pt}) can be written as
%
\begin{align}
c(u,v) = \frac{2}{\pi^2}
\biggl[
&\bar{D}_{2222}\left(1  -6v - \frac{v}{u} - \frac{v^2}{u}+ v^2\right)
+\bar{D}_{1212}\left( \frac{v}{u}+\frac{v^2}{u}\right) \nonumber \\
& +\left( 1-v\right) (\bar{D}_{2211}-\bar{D}_{2112})
+4v(\bar{D}_{3322} +\bar{D}_{3223})\biggl] \ ,
\end{align}
where $\bar{D}$ is expressed in the integral form as
\begin{align}
\bar{D}_{\Delta_1\Delta_2\Delta_3\Delta_4}=
 2K\int_0^\infty &{\rm d} t_1 \cdots {\rm d} t_4
t_1^{\Delta_1-1}t_2^{\Delta_2-1}t_3^{\Delta_3-1}t_4^{\Delta_4-1}
\nonumber \\
&\times {\rm  exp}\Biggl[-t_1t_2-t_1t_3-t_1t_4-t_2t_3-
\frac{v}{u}t_2t_4-v t_3t_4\Biggl] \ ,
\label{dbar} \\
\text{with} &~~~~ K\,=\,\frac{\pi^{\frac{d}{2}}\Gamma(\frac{
    \Delta_1+\Delta_2+\Delta_3+\Delta_4}{2}-\frac{d}{2})}{
  2\Gamma(\Delta_1)\cdots\Gamma(\Delta_4)} \ . \nonumber
\end{align}

For numerical evaluation, it is convenient to express $\bar{D}$
in the form of infinite series in $v$ and $Y=1-\frac{v}{u}$ as
\begin{align}
\bar{D}_{2222}(v,Y)=&\pi^2\sum_{n,m=0}^{\infty}\frac{Y^m}{m!}\frac{v^n}{(n!)^2}
 \frac{\Gamma(n+2)^2\Gamma(2+n+m)^2}{\Gamma(4+2n+m)}\nonumber \\
&\times  \left(
-\frac{1}{n+1}+\psi(4+2n+m)-\psi(n+m+2)-\frac{1}{2}\ln{v}
\right) \ ,  \nonumber \\
\bar{D}_{1212}(v,Y)=&\pi^2
\sum_{n,m=0}^{\infty}\frac{Y^m}{m!}\frac{v^n}{(n!)^2}
 \frac{\Gamma(n+1)^2\Gamma(n+m+2)^2}{\Gamma(3+2n+m)}\nonumber \\
&\times  \left(\psi(3+2n+m)-\psi(n+m+2)-\frac{1}{2}\ln{v}
\right)\ ,   \nonumber \\
\bar{D}_{2211}(v,Y)=&-\frac{\pi^2}{2}
\sum_{n,m=0}^{\infty}\frac{Y^m}{m!}\frac{v^n}{(n!)^2}
 \frac{n\Gamma(n+1)^2\Gamma(n+m+1)^2}{\Gamma(2+2n+m)}\nonumber \\
&\times  \left(-\frac{1}{n}-2\psi(n+m+1)+2\psi(2+2n+m)-\ln{v}
\right)\ , \nonumber \\
\bar{D}_{2112}(v,Y)=&\frac{\pi^2}{2}
\sum_{n,m=0}^{\infty}\frac{Y^m}{m!}\frac{v^n}{(n!)^2}
\frac{\Gamma(n+2)\Gamma(n+1)\Gamma(n+m+1)\Gamma(n+m+2)}{\Gamma(3+2n+m)}
\nonumber \\
&\times  \left(
-\frac{1}{n+1}+2\psi(3+2n+m)-\psi(n+m+1)-\psi(n+m+2)-\ln{v}
\right)\ , \nonumber \\
\bar{D}_{3322}(v,Y)=&-\frac{\pi^2}{4}
\sum_{n,m=0}^{\infty}\frac{Y^m}{m!}\frac{v^n}{(n!)^2}
 \frac{n\Gamma(n+2)^2\Gamma(2+n+m)^2}{\Gamma(4+2n+m)}
\nonumber \\
&\times  \left(-\frac{3n+1}{n(n+1)}+2\psi(4+2n+m)-2\psi(2+n+m)-\ln{v}
 \right) \ ,  \nonumber \\
\bar{D}_{3223}(v,Y)=&\frac{\pi^2}{4}
\sum_{n,m=0}^{\infty}\frac{Y^m}{m!}\frac{v^n}{(n!)^2}
\frac{\Gamma(n+2)\Gamma(n+3)\Gamma(2+n+m)\Gamma(3+n+m)}{\Gamma(5+2n+m)}
\nonumber \\
&\times  \left(-\frac{3n+5}{(n+1)(n+2)}+2\psi(5+2n+m)\right. \nonumber \\
&~~~~~-\psi(2+n+m)-\psi(3+n+m)-\ln{v}
 \bigg) \ .
\label{dbar-u}
\end{align}
The parameters
$u$ and $v$ are positive by definition (\ref{cross-ratios}).
These expressions converge fast in the region $v<1$ and $Y>0$
(the latter inequality corresponds to $u>v$).
Note that, for $Y<0$, each term of the series in $Y$ changes its sign,
which makes numerical evaluation hard.

In order to obtain the expressions suitable
for numerical evaluation in the other regions of $u$ and $v$,
we consider a change of integration variables
in the integral form (\ref{dbar}) of $\bar{D}$,
\begin{align}
t_1^\prime = t_3  \ , \quad
t_2^\prime = t_2 \ , \quad
t_3^\prime = t_1 \ , \quad
t_4^\prime = v t_4  \ ,
\end{align}
which gives
\begin{align}
\bar{D}_{\Delta_1\Delta_2\Delta_3\Delta_4}
\left(v,1-\frac{v}{u}\right)=
& \frac{2K}{v^{\Delta_4}}
\int_0^\infty {\rm d} t_1^\prime  \cdots {\rm d} t_4^\prime
t_1^{\prime \Delta_3-1}
t_2^{\prime \Delta_2-1}
t_3^{\prime \Delta_1-1}
t_4^{\prime \Delta_4-1}
\nonumber \\
&~~~~~~~
\times {\rm exp}
\Biggl[-t_1^\prime t_2^\prime -t_1^\prime t_3^\prime -t_1^\prime t_4^\prime
-t_2^\prime t_3^\prime -
\frac{1}{u}t_2^\prime t_4^\prime -\frac{1}{v} t_3^\prime t_4^\prime
\Biggl]
\nonumber \\
=& \frac{1}{v^{\Delta_4}}
\bar{D}_{\Delta_3\Delta_2\Delta_1\Delta_4}
\left(\frac{1}{v},1-\frac{1}{u}\right)
\ .
\label{dbar-uv}
\end{align}
By using the infinite series (\ref{dbar-u}) for the
$\bar{D}$ on the right-hand side of (\ref{dbar-uv}),
we can evaluate the left-hand side in the region $u>1$ and $v>1$.

Another change of integration variables
\begin{align}
t_1^\prime = t_2 \ , \quad
t_2^\prime = t_1 \ , \quad
t_3^\prime = t_3 \ , \quad
t_4^\prime = \frac{v}{u} t_4 \ ,
\end{align}
leads to
\begin{align}
\bar{D}_{\Delta_1\Delta_2\Delta_3\Delta_4}
\left( v,1-\frac{v}{u} \right)=
& 2K \bigg(\frac{u}{v}\bigg)^{\Delta_4}
\int_0^\infty {\rm d} t_1^\prime  \cdots {\rm d} t_4^\prime
t_1^{\prime \Delta_2-1}t_2^{\prime \Delta_1-1}t_3^{\prime \Delta_3-1}t_4^{\prime \Delta_4-1}
\nonumber \\
&~~~~~~~
\times \exp
\Biggl[-t_1^\prime t_2^\prime -t_1^\prime t_3^\prime -t_1^\prime t_4^\prime
-t_2^\prime t_3^\prime -
\frac{u}{v}t_2^\prime t_4^\prime -u t_3^\prime t_4^\prime \Biggl]  \nonumber \\
=& \bigg(\frac{u}{v}\bigg)^{\Delta_4}
 \bar{D}_{\Delta_2\Delta_1\Delta_3\Delta_4}
\left( u,1-\frac{u}{v} \right) \ .
\label{dbar-v}
\end{align}
By using the infinite series (\ref{dbar-u}) for the
$\bar{D}$ on the right-hand side of (\ref{dbar-v}),
we can evaluate the left-hand side in the region $u<1$ and $v>u$.


\section{Numerical evaluation of the predicted four-point function}
\label{appendix:B}

The prediction of the AdS/CFT correspondence
for the four-point function (\ref{4-point function})
is given in terms of the function $c(u,v)$
in (\ref{ads-cft-pred-4pt}).
In order to translate this prediction
into the form that can be compared
with our Monte Carlo results (\ref{G4-def}),
we have to perform
the integral in (\ref{4pt-measured}), which we do numerically
in the following way.

In terms of the coordinates on $R^4$,
the integral (\ref{4pt-measured})
can be rewritten as
\begin{align}
I(p_1, p_2 , p_3 , p_4)
= \int \left(\prod_{i=1}^4 d^4x_i  |x_i|^{-2-ip_i}  \right)
\frac{c(u,v)}{x_{12}^2 x_{23}^2 x_{34}^2 x_{41}^2} \ ,
\label{integral}
\end{align}
where $|x_i|=\sqrt{\sum_{\mu=1}^4x_i^{\mu}x_i^{\mu}}$
and we have omitted
a numerical factor $\zeta^2 \lambda^4 /(2^8 \pi^8 N^2)$
in (\ref{4pt-measured}).
We regard (\ref{integral}) as an expectation value of
$c(u,v) \prod_{i=1}^4|x_i|^{-ip_i}$ with respect to the
partition function
\begin{align}
Z= \int \left(\prod_{i=1}^4 d^4x_i  \frac{1}{|x_i|^{2}}  \right)
\frac{1}{x_{12}^2 x_{23}^2 x_{34}^2 x_{41}^2} \ .
\label{pf}
\end{align}
Note that the system (\ref{pf}) has symmetries
under the inversion $x_i \rightarrow x_i/|x_i|^2$,
the dilatation $x_i \rightarrow qx_i$ and the rotation of
the four-dimensional vectors $x_i$.
We fix these symmetries as follows.
First we insert $1= \int dR \; \delta (R-|x_1|)$ in the integrand
to fix the dilatation symmetry.
After the insertion, we rescale the
integration variables as $x_i \rightarrow R x_i$.
Then the integration over $R$ factorizes and yields
a delta function representing
the momentum conservation that appears on the left-hand side
of (\ref{4pt-measured}).
In the remaining integral, $x_1$ can be set to $(1,0,0,0)$
using the delta function and the rotational symmetry.
Finally, using the inversion symmetry, we restrict
the integration over $x_2$, $x_3$ and $x_4$ to the region
in which two of them have norms less than 1.

Then the function $F(p_1,p_2,p_3,p_4)$ in the ratio (\ref{RATIO4pt})
can be evaluated by
\begin{align}
F(p_1,p_2,p_3,p_4)
= \frac{\left\langle c(u,v) \prod_{i=1}^4|x_i|^{-ip_i} \right\rangle }
{\left\langle \prod_{i=1}^4|x_i|^{-ip_i} \right\rangle} \ ,
\label{ratio1}
\end{align}
where the expectation value
$\langle \cdots \rangle $ is
taken with respect to the partition function (\ref{pf})
after fixing the symmetries as described above.
We apply the standard Hybrid Monte Carlo algorithm to calculate
the expectation values.
The function $c(u,v)$ is evaluated by using
the infinite series in appendix \ref{appendix:A}
truncated at sufficiently high order.
Figure~\ref{figures-4pt_2} (Right) shows the results for (\ref{ratio1})
obtained in this way.
Note that the observables in (\ref{ratio1}) involve
a phase factor $|x_i|^{-ip_i}$,
which makes both the numerator and the denominator
in (\ref{ratio1}) exponentially small as $p_i$ increases.
For this reason, we were able to calculate $F(p_1,p_2,p_3,p_4)$
for momentum
configurations $(p,-p,p,-p)$ and $(p,p,-p,-p)$
only for $p\le 4$ within statistical errors of a few \%.



\section{Fuzzy spherical harmonics}
\label{appendix:E}

In this section we present some formulae
for fuzzy spherical harmonics used in
section \ref{sec:corresponding-PWMM}.
The fuzzy spherical harmonics $Y^{(n,n')}_{jm}$ is
a basis for $n\times n'$ matrices
defined by
\begin{align}
Y^{(n,n')}_{jm}=\sum_{r=-p}^p\sum_{r'=-p'}^{p'}(-1)^{-p+r'}C^{jm}_{pr\:p'r'}
|pr\rangle\langle p'r'| \ ,
\label{definition of fuzzy spherical harmonics}
\end{align}
where $p=(n-1)/2,\;\;p'=(n'-1)/2$.
The symbol $|pr\rangle$ represents the basis of the
spin $p$ representation of ${\rm SU}(2)$,
and $C^{jm}_{pr\:p'r'}$ denotes the Clebsch-Gordan coefficient.
The details of the fuzzy spherical harmonics
can be found, for instance,
in refs.~\cite{Ishiki:2006yr,Ishii:2008tm,Ishii:2008ib}.
It is easy to derive the following relations
\begin{align}
&L^{(n)}_{\pm}Y_{jm}^{(n,n')}-Y_{jm}^{(n,n')}L^{(n')}_{\pm}
=\sqrt{(j\mp m)(j\pm m+1)}Y_{jm\pm 1}^{(n,n')} \ ,  \nonumber\\
&L^{(n)}_{3}Y_{jm}^{(n,n')}-Y_{jm}^{(n,n')}L^{(n')}_{3}
=mY_{jm}^{(n,n')} \ , \nonumber\\
& \Bigl(Y_{jm}^{(n,n')}\Bigr)^{\dagger}
=(-1)^{m-\frac{1}{2}(n-n')}Y_{j-m}^{(n',n)} \ , \nonumber\\
&\mbox{tr} \Bigr(Y^{(n,n')}_{jm}Y^{(n',n)}_{j'm'}\Bigl)=
(-1)^{m-\frac{1}{2}(n-n')}\delta_{jj'}\delta_{m\:-m'} \ ,
\label{property of fuzzy spherical harmonics}
\end{align}
where tr represents
the trace over $n\times n$ matrices.

\section{Free theory results with finite regularization parameters}
\label{appendix:F}

In this section we present free theory results
for the correlation functions
with finite regularization parameters $\beta, \Lambda, n, \nu, k$.
(The first two parameters represent the IR and UV cutoffs in the
$t$-direction, and the latter three parameters appear
in the background (\ref{our background}).)
These results are used in normalizing our Monte Carlo data
in eq.~(\ref{ratio-2pt}), (\ref{ratio-3pt}) and
(\ref{ratio-4pt}). They can be derived in the way described in
section \ref{sec:corresponding-PWMM}.

The two-point function for the free theory is given by
\begin{equation}
 \left\langle
 \widetilde{O}_{45}(p) \,  \widetilde{O}_{54}(-p) \right\rangle_{\rm PW,free}
= \frac{\beta^2 k^2}{(2\pi )^4}
\sum_{I,J=1}^\nu \sum_{j=\frac{1}{2}|n_I -n_J|}^{\frac{1}{2}(n_I +n_J )-1} \sum_{n=p-\Lambda}^{\Lambda}
   \frac{2j+1}{(n^2 +b_j^2 )\{ (n-p)^2 +b_j^2 \} } \ ,
\end{equation}
where
\begin{equation}
b_j = \frac{\beta \mu (2j+1)}{4\pi}  \ .
\end{equation}
The three-point function we consider in this work is given for
the free theory by
\begin{equation}
 \left\langle
 \widetilde{O}_{45}(p) \,  \widetilde{O}_{56}(0)  \widetilde{O}_{64}(-p) \right\rangle_{\rm PW,free}
= \frac{\beta^3 k^2}{(2\pi )^6} \sum_{I,J=1}^\nu \sum_{j=\frac{1}{2}|n_I -n_J|}^{\frac{1}{2}(n_I +n_J )-1}  \sum_{n=p-\Lambda}^{\Lambda}
   \frac{2j+1}{(n^2 +b_j^2 )^2 \{ (n-p)^2 +b_j^2 \} } \ .
\end{equation}
We consider four-point functions with three different
types of momentum configuration,
which are given for the free theory, respectively, as
\begin{eqnarray}
&& \left\langle
 \widetilde{O}_{45}(p) \,  \widetilde{O}_{56}(0)
\widetilde{O}_{67}(0)  \widetilde{O}_{74}(-p)
\right\rangle_{\rm PW,free} \nonumber\\
&&= \frac{\beta^4 k^2}{(2\pi )^8}
\sum_{I,J=1}^\nu
\sum_{j=\frac{1}{2}|n_I -n_J|}^{\frac{1}{2}(n_I +n_J )-1}
\sum_{n=p-\Lambda}^{\Lambda}
   \frac{2j+1}{(n^2 +b_j^2 )^3 \{(n-p)^2 +b_j^2 \} } \ , \nonumber\\
&& \left\langle
 \widetilde{O}_{45}(p) \,  \widetilde{O}_{56}(-p)
\widetilde{O}_{67}(p)  \widetilde{O}_{74}(-p)
\right\rangle_{\rm PW,free} \nonumber\\
&&= \frac{\beta^4 k^2}{(2\pi )^8} \sum_{I,J=1}^\nu
\sum_{j=\frac{1}{2}|n_I -n_J|}^{\frac{1}{2}(n_I +n_J )-1}
\sum_{n=p-\Lambda}^{\Lambda}
   \frac{2j+1}{(n^2 +b_j^2 )^2 \{ (n-p)^2 +b_j^2 \}^2} \ , \nonumber\\
&& \left\langle
 \widetilde{O}_{45}(p) \,  \widetilde{O}_{56}(p)
\widetilde{O}_{67}(-p)  \widetilde{O}_{74}(-p)
\right\rangle_{\rm PW,free} \nonumber\\
&&= \frac{\beta^4 k^2}{(2\pi )^8}
\sum_{I,J=1}^\nu
\sum_{j=\frac{1}{2}|n_I -n_J|}^{\frac{1}{2}(n_I +n_J )-1}
\sum_{n=p-\Lambda}^{-p+\Lambda }
   \frac{2j+1}{(n^2 +b_j^2 )^2 \{(n-p)^2 +b_j^2 \} \{(n+p)^2 +b_j^2 \} } \ .
\end{eqnarray}

\section{Stability of the background}
\label{appendix:C}

\begin{table}[t]
\begin{center}
\begin{tabular}[b]{|c|c|c|}
\hline
\mbox{vacuum} &
$X_{i}/\mu$     &$(m_{1},m_{2},m_{3},m_{4},m_{5},m_{6})$  \\ \hline \hline
(a) &
$ L_i^{(1)} \otimes {\bf 1}_{6}$    &$\left( 0,0,0,0,0,0 \right)$ \\ \hline
(b) &
$ L_i^{(1)} \otimes {\bf 1}_{4}\oplus L_i^{(2)} \otimes {\bf 1}_{1} $
&$\left( -\frac{1}{2},0,0,0,0,\frac{1}{2} \right)$ \\ \hline
(c) &
$\bigl( L_i^{(1)}  \oplus  L_i^{(2)}  \bigr) \otimes {\bf 1}_{2}$
&$\left( -\frac{1}{2},-\frac{1}{2},0,0,\frac{1}{2},\frac{1}{2}\right)$ \\ \hline
(d) &
$ L_i^{(2)} \otimes {\bf 1}_{3}$
&$\left( -\frac{1}{2},-\frac{1}{2},
-\frac{1}{2},\frac{1}{2},\frac{1}{2},\frac{1}{2} \right)$ \\ \hline
(e) &
$ L_i^{(1)} \otimes {\bf 1}_{3}\oplus L_i^{(3)} \otimes {\bf 1}_{1} $
&$\left( -1,0,0,0,0,1 \right)$ \\ \hline
(f) &
$\bigl( L_i^{(1)}  \oplus  L_i^{(2)}  \oplus  L_i^{(3)} \bigr)
\otimes {\bf 1}_{1}$
&$\left( -1,-\frac{1}{2},0,0,\frac{1}{2},1 \right)$ \\ \hline
(g) &
$ L_i^{(3)} \otimes {\bf 1}_{2}$
&$\left( -1,-1,0,0,1,1 \right)$ \\ \hline
(h) &
$ L_i^{(1)} \otimes {\bf 1}_{2}\oplus L_i^{(4)} \otimes {\bf 1}_{1} $
&$\left( -\frac{3}{2},-\frac{1}{2},0,0,\frac{1}{2},\frac{3}{2} \right)$
\\ \hline
(i) &
$\bigl( L_i^{(2)}  \oplus  L_i^{(4)} \bigr) \otimes {\bf 1}_{1}$
&$\left( -\frac{3}{2}-\frac{1}{2},-\frac{1}{2},
\frac{1}{2},\frac{1}{2},\frac{3}{2} \right)$ \\ \hline
(j) &
$\bigl( L_i^{(1)}  \oplus  L_i^{(5)} \bigr) \otimes {\bf 1}_{1}$
&$\left( -2,-1,0,0,1,2 \right)$ \\ \hline
(k) &
$ L_i^{(6)} \otimes {\bf 1}_{1}$
&$\left( -\frac{5}{2},-\frac{3}{2},-\frac{1}{2},
\frac{1}{2},\frac{3}{2},\frac{5}{2} \right)$ \\
\hline
\end{tabular}
\end{center}
\caption{The list of all possible 11 classical vacua of the PWMM
with $N=6$. The right column shows
the corresponding eigenvalues $m_{p}\ (p=1,\cdots ,6)$
of $(X_{i}/\mu)$ for each vacuum.}
\label{possible_vac}
\end{table}

In our simulation,
we start from a classical vacuum (\ref{our background}) of the PWMM
with the parameters $(n,\nu ,k)=(\frac{3}{2},2,2)$,
which corresponds to the background
%
\begin{align}
X_i=\mu \Bigl( L_i^{(1)}  \oplus  L_i^{(2)}  \Bigr) \otimes {\bf
  1}_{2}
\quad
\mbox{for $i=1,2,3$} \ .
\label{our_bg}
\end{align}
For the validity of the large-$N$ reduction,
we have to make sure that the configurations generated by Monte Carlo
simulation fluctuate around (\ref{our_bg})
and do not make a transition to other vacua.
Such a transition is suppressed in the large-$N$ limit (\ref{limit})
for arbitrary coupling constant
as we discussed at the end of section \ref{sec:SYMrxs3-from-pwmm},
but it can occur for finite $N$ at sufficiently strong coupling.

In order to probe the possible transitions to other vacua,
we consider the eigenvalues $m_{p}\ (p=1,\cdots ,6)$
of $(X_{i}/\mu)$ for each of $i=1,2,3$
with the ordering $m_1 \le m_2 \le \ldots \le m_6$.
(The eigenvalue distribution is the same for $i=1,2,3$
for the classical vacua due to the SO(3) symmetry.)
For instance, our background (\ref{our_bg}) gives
\begin{equation}
(m_{1},m_{2},m_{3},m_{4},m_{5},m_{6})
= \left( -\frac{1}{2},-\frac{1}{2},0,0,\frac{1}{2},\frac{1}{2}\right) \ .
\label{classical_eigen}
\end{equation}
In table \ref{possible_vac},
we list all the classical vacua in the PWMM for the matrix size
$N=6$ and the corresponding eigenvalues $m_{p}\ (p=1,\cdots ,6)$.
In Monte Carlo simulation,
the eigenvalues $m_{p}$ fluctuate around (\ref{classical_eigen})
in the weakly coupled (small $g_{\rm PW}$) regime,
but in the strongly coupled (large $g_{\rm PW}$) regime,
the system may undergo a transition to a different vacuum,
which can be seen as a change of the eigenvalue distribution
from (\ref{classical_eigen}) into another one in table \ref{possible_vac}.

\begin{figure}[t]
\begin{center}
\includegraphics[width=74mm]{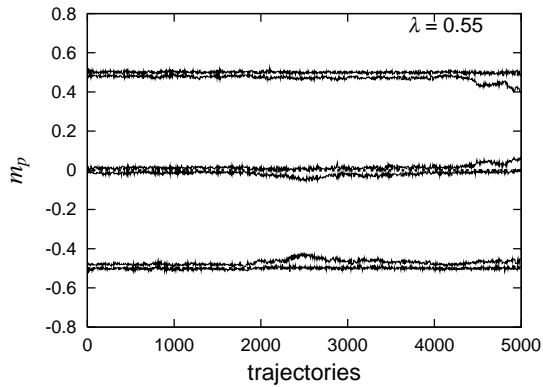}
\end{center}
\caption{The history of the eigenvalues $m_{p}$
after thermalization is plotted for $\lambda=0.55$.
}
\label{history}
\end{figure}

\begin{figure}[t]
\begin{center}
\includegraphics[width=49mm]{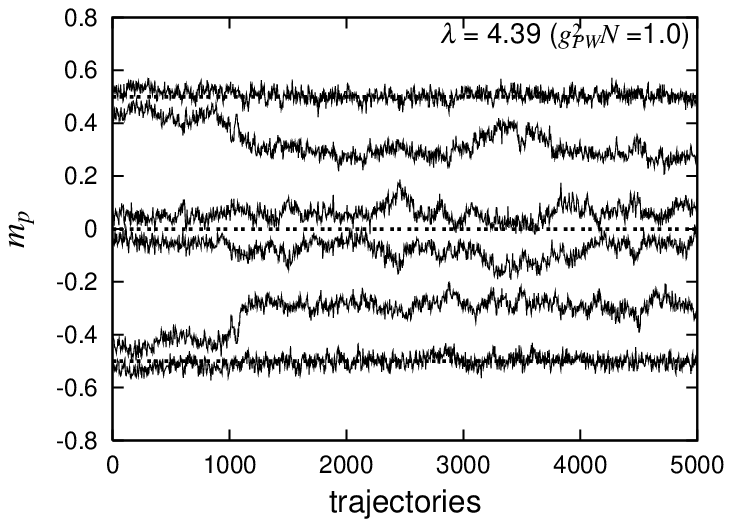}
\includegraphics[width=49mm]{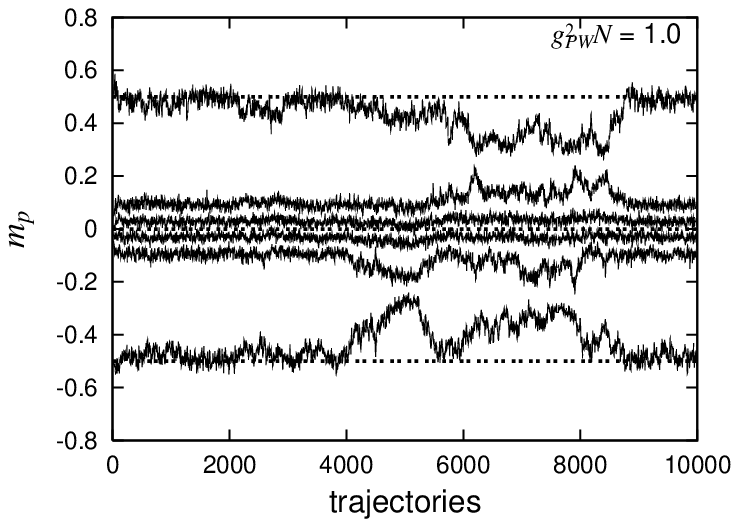}
\includegraphics[width=49mm]{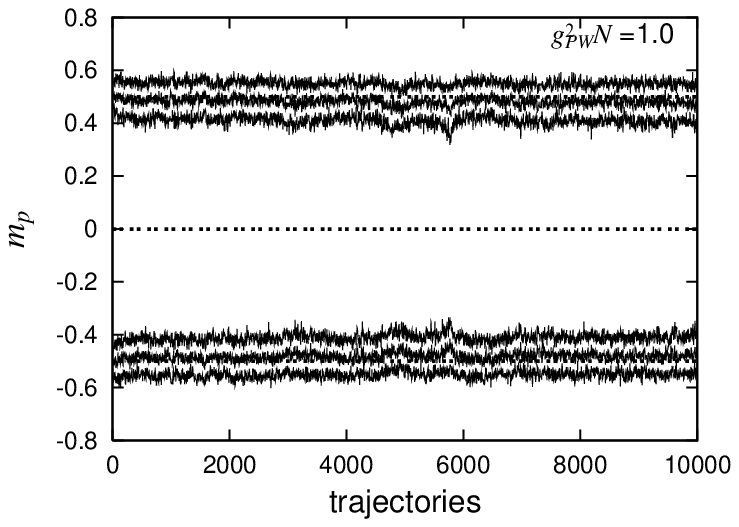}\\
\includegraphics[width=49mm]{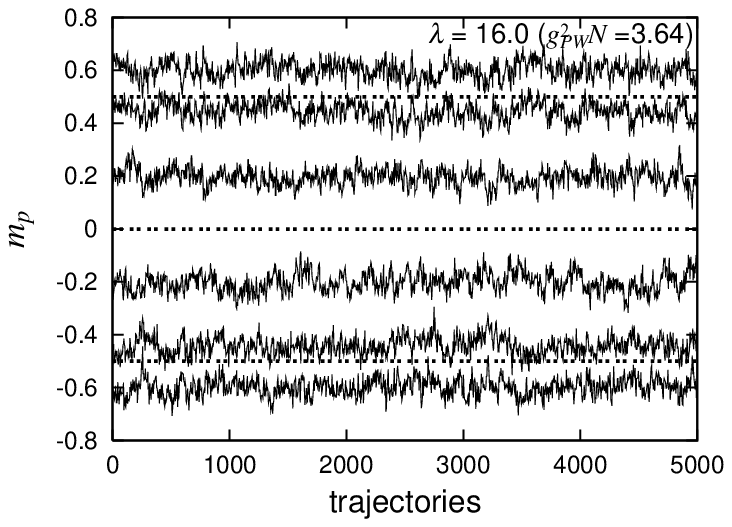}
\includegraphics[width=49mm]{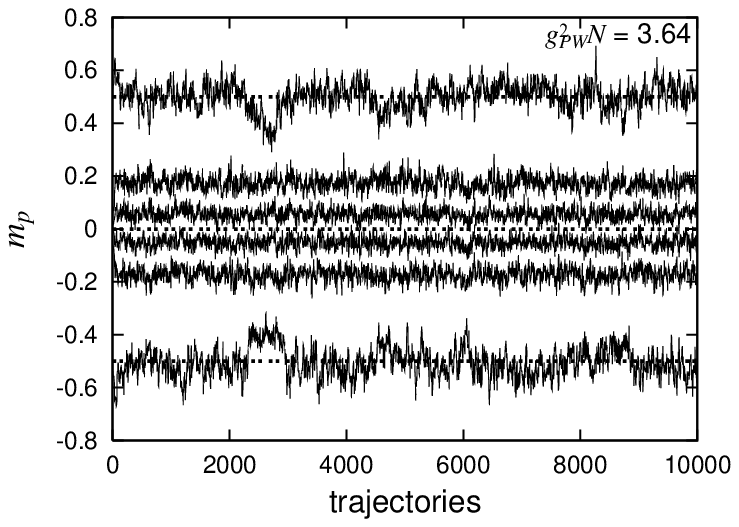}
\includegraphics[width=49mm]{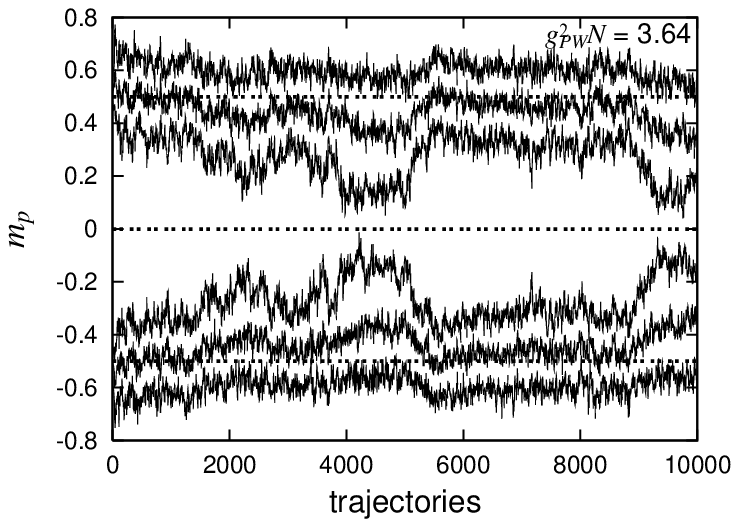}
\end{center}
\caption{The history of the eigenvalues $m_p$
after thermalization for three different
initial configurations.
The left column is for our background (\ref{our_bg}),
namely the vacuum (c) in table \ref{possible_vac}.
The middle column and the right column are for
the vacua (b) and (d) in table \ref{possible_vac}, respectively.
The plots on the top and the bottom are the
results for the coupling constant $g_{\rm PW}^{2}N =1.0,3.64$,
which correspond to $\lambda= 4.39,16.0$, respectively,
in the case of our background (\ref{our_bg}).
}
\label{other_bg}
\end{figure}

Figure \ref{history} shows the history of the eigenvalues $m_{p}$
after thermalization
in the weak coupling case $\lambda = 0.55$.
Here we take the mean value with respect to $i=1,2,3$
for each of $m_p$.
We find that the eigenvalues $m_{p}$
fluctuate around the classical values (\ref{classical_eigen}).
We plot the results for $\lambda= 4.39,16.0$
in the left column of figure~\ref{other_bg}.
We observe considerable deviation
from the classical values (\ref{classical_eigen}),
which are represented by the horizontal lines.

Note, in particular, that there are
three classical vacua (b), (c) and (d) in table \ref{possible_vac},
which has the largest eigenvalue $m_6=\frac{1}{2}$.
In order to make sure that transitions among these vacua
do not occur,
we perform simulations starting from
the vacua (b) and (d)
with the coupling constant $g_{\rm PW}^{2}N =1.0,3.64$,
which corresponds to $\lambda= 4.39,16.0$
due to eq.~(\ref{g-lambda-rel})
in the case of our background (\ref{our_bg}).
%
The histories of the eigenvalues $m_{p}$
for the two initial configurations
are shown
in the middle column and the right column of figure~\ref{other_bg},
respectively.
We find that the histories can be clearly distinguished from
the one in the left column.
Thus we conclude that our background (\ref{our_bg})
remains stable up to $\lambda= 16.0$.

\begin{figure}[t]
\begin{center}
\includegraphics[width=72mm]{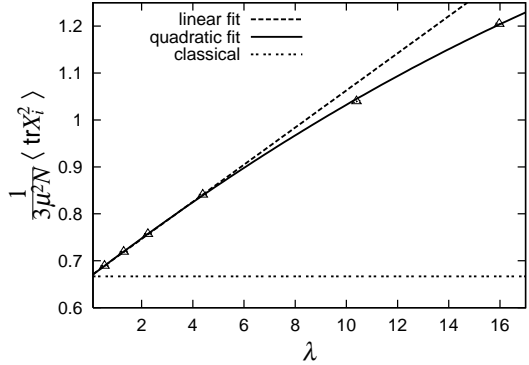}
\end{center}
\caption{The expectation value
(\ref{casimir-def})
obtained in simulations starting from our background (\ref{our_bg})
is plotted against $\lambda$.
The dotted line represents the classical value.
The dashed line and the solid line
represent the fits to linear and quadratic behaviors in $\lambda$,
respectively.
}
\label{trxt}
\end{figure}

In order to confirm this conclusion further,
we calculate
\begin{align}
\frac{1}{3\mu^{2}\beta N}
\left\langle \int_0^\beta dt \,
{\rm tr} X_{i}(t)^{2} \right\rangle
\label{casimir-def}
\end{align}
for various $\lambda$
by starting simulations from our background (\ref{our_bg}).
The results are plotted against $\lambda$ in figure~\ref{trxt}.
The data points can be nicely fitted to a quadratic behavior
up to $\lambda= 16.0$.
This implies that no phase transition occurs within this region,
which is consistent with our conclusion from figure \ref{other_bg}.

\begin{figure}[t]
\begin{center}
\includegraphics[width=72mm]{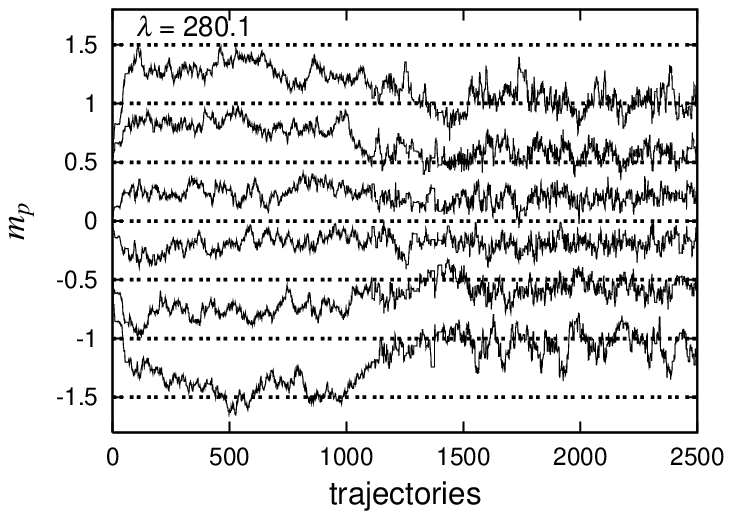}
\includegraphics[width=72mm]{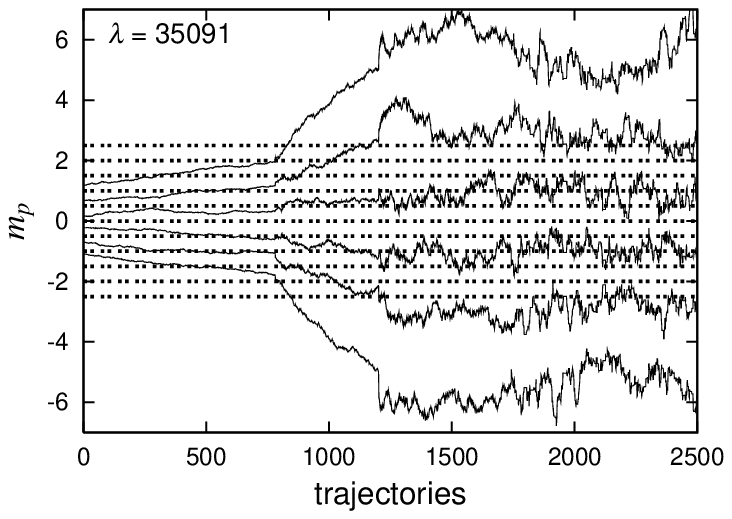}
\end{center}
\caption{The history of the eigenvalues of $m_{p}$
at very strong coupling $g_{\rm PW}^{2}N = 63.9,~8000$,
which would have corresponded to $\lambda= 280.1,35091$
if the background (\ref{our_bg}) were unbroken.
The horizontal lines represent
$m_{p}=\pm\frac{5}{2},\pm 2,\pm\frac{3}{2},\pm 1,\pm\frac{1}{2},0$,
which correspond to possible values $m_{p}$ for classical vacua
in the PWMM for $N=6$.
}
\label{transition}
\end{figure}

In figure~\ref{transition}
we show some results
starting from our background (\ref{our_bg})
at very strong couplings $g_{\rm PW}^{2}N = 63.9$ (Left)
and $g_{\rm PW}^{2}N = 8000$ (Right),
which would have corresponded to $\lambda= 280.1$, $35091$
in SYM, respectively,
if our background (\ref{our_bg}) were unbroken.
The plot in the left panel suggests the occurrence of
a transition from our background (\ref{our_bg}) to
the vacuum (f)
after 1500 trajectories.
The plot in the right panel
suggests that the obtained configurations cannot be
understood by small fluctuations from one of the classical vacua
since the eigenvalue distribution exceeds
the largest possible value of $|m_p| =\frac{5}{2}$ for the classical vacua
in the PWMM with $N=6$.


\section{Dependence on regularization parameters}
\label{appendix:D}

In this section we discuss the dependence
of our results
on the regularization parameters.
In particular, we have used $\Lambda=12$ and the background
$(n,\nu , k ) = (\frac{3}{2},2 , 2)$.
We change one of $\Lambda$, $k$ and $(n,\nu)$ fixing the others,
and study how our results
for the ratio $R^{\rm (2)}(p,-p)$ of the two-point functions
in (\ref{ratio-2pt}) are affected
for $\lambda=16.0$ and $\beta=3.25$.
As for the choice of $\beta$, see footnote \ref{footnote:beta-choice}.


Let us first consider the dependence on
the parameter $\Lambda$, which plays the role
of a UV cutoff in the $t$-direction.
Since 16 out of 32 SUSYs of $\mathcal{N}=4$ SYM
restore in the $\Lambda\rightarrow\infty$ limit,
it is important whether the value $\Lambda =12$ we have chosen
is large enough.

\begin{figure}[t]
\begin{center}
\includegraphics[width=72mm]{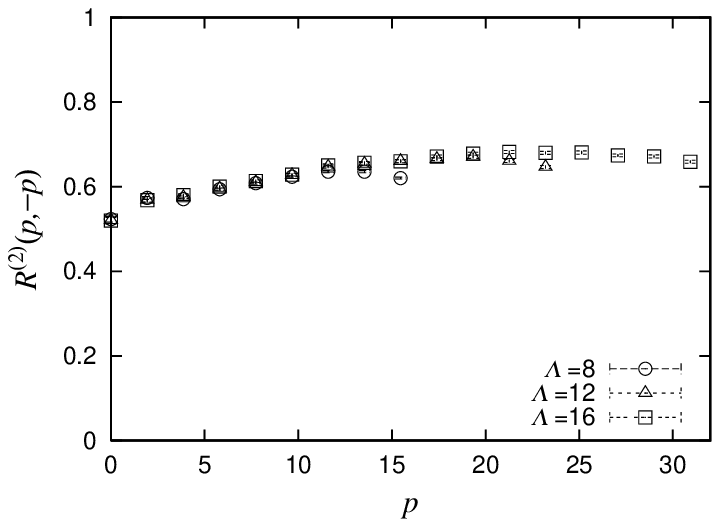}
\includegraphics[width=72mm]{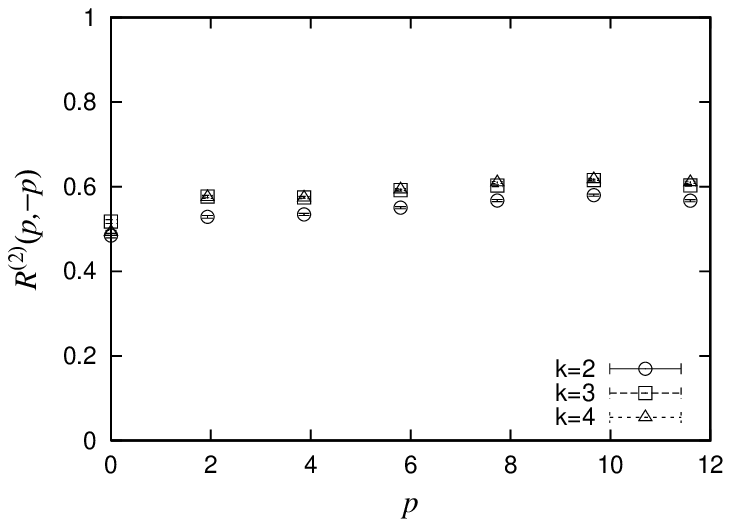}
\end{center}
\caption{(Left) The ratio $R^{\rm (2)}(p,-p)$
of the two-point functions for $\lambda=16.0$ is plotted
against $p$
for various values of $\Lambda$
with $(n,\nu ,k)= (\frac{3}{2},2,2)$.
(Right) The ratio $R^{\rm (2)}(p,-p)$
of the two-point functions for $\lambda=16.0$
is plotted against $p$
for $k=2,3,4$ with $(n,\nu)= (\frac{3}{2},2)$
and $\Lambda=6$.
}
\label{lam-k-dependence}
\end{figure}

In figure~\ref{lam-k-dependence} (Left)
we plot the ratio $R^{\rm (2)}(p,-p)$ of the two-point functions
in (\ref{ratio-2pt})
against $p$ for various $\Lambda$
with $(n,\nu , k ) = (\frac{3}{2},2 , 2)$.
This confirms that the finite-$\Lambda$ effects
for our choice $\Lambda=12$
are negligible
in the momentum region $p=\frac{2\pi}{\beta} n$ with $n\le 10$.



The parameter $k$ represents the number of coincident
fuzzy spheres with each radius,
and it corresponds to the rank of the gauge group in
${\cal N}=4$ SYM.
The large-$k$ limit should be taken to make sure that planar diagrams
dominate and to suppress the transition to other classical vacua.
Since all the fields in the PWMM (or in $\mathcal{N}=4$ SYM)
are in the adjoint representation,
it is expected that the finite-$k$ effects are of
the order of $\mathcal{O}(1/k^{2})$.
Figure \ref{lam-k-dependence} (Right)
shows the ratio $R^{\rm (2)}(p,-p)$
as a function of $p$ for $k=2,3,4$
with $\Lambda=6$ and $(n,\nu) = (\frac{3}{2},2)$.
Indeed we only find little dependence on $k$.

\begin{figure}[t]
\begin{center}
\includegraphics[width=72mm]{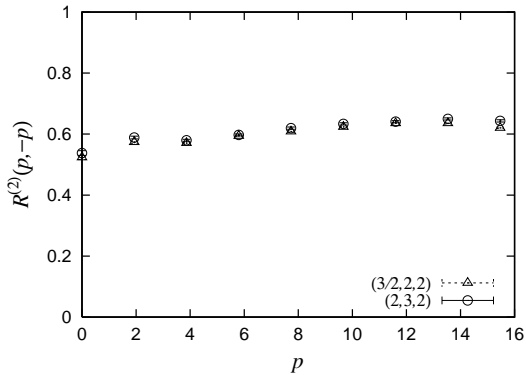}
\end{center}
\caption{The ratio $R^{\rm (2)}(p,-p)$
of the two-point functions for $\lambda=16.0$
is plotted against $p$
for the two backgrounds
$(n, \nu, k) =(\frac{3}{2},2, 2)$
and $(n, \nu, k) =(2,3,2)$ with $\Lambda=8$.
}
\label{R_bg}
\end{figure}

Finally we study the dependence on
the parameters $n$ and $\nu$ in the
background (\ref{our background}).
They play the role of UV cutoffs on $S^{3}$,
and the limits $n,\nu \rightarrow\infty$
in (\ref{limit}) are important, in particular, for
the full superconformal symmetry to be restored.
Here we compare the results for our $N=6$ background
$(n, \nu, k) =(\frac{3}{2},2, 2)$,
which corresponds to (\ref{our_bg}),
with those for the $N=12$ background
$(n, \nu, k) =(2,3,2)$, which corresponds to
\begin{align}
X_i=\mu \bigl( L_i^{(1)} \oplus L_i^{(2)}
\oplus L_i^{(3)} \bigr) \otimes {\bf 1}_{2}
\quad
\mbox{for $i=1,2,3$} \ .
\label{larger_bg}
\end{align}
Figure \ref{R_bg}
shows the plot of $R^{\rm (2)}(p,-p)$ as a function of $p$
for the two backgrounds with $\Lambda=8$.
We find that the results for (\ref{larger_bg}) increases slightly
compared with the results for our background (\ref{our_bg}).
This is consistent with our speculation that
$R^{\rm (2)}(p,-p)$ approaches 1 in the limit (\ref{limit})
for arbitrary coupling $\lambda$.





\end{document}